\documentclass[letterpaper,english,reprint,aps,nofootinbib]{revtex4-1}
\usepackage[T1]{fontenc}
\usepackage[latin9]{inputenc}
\usepackage{babel}
\usepackage{amsmath}
\usepackage{amssymb}
\usepackage{graphicx}
\usepackage{color}
\usepackage{float}
\usepackage{longtable}
\usepackage[unicode=true,pdfusetitle,
bookmarks=true,bookmarksnumbered=false,bookmarksopen=false,
breaklinks=false,pdfborder={0 0 1},backref=false,colorlinks=false]
{hyperref}

\makeatletter

\pdfpageheight\paperheight
\pdfpagewidth\paperwidth

\providecommand{\tabularnewline}{\\}

\@ifundefined{showcaptionsetup}{}{%
\PassOptionsToPackage{caption=false}{subfig}}
\usepackage[caption=false]{subfig}
\makeatother

\begin{document}
\title{Ratio of strange to $u/d$ momentum fraction in disconnected insertions}
\author{Jian Liang$^{1}$\footnote[1]{jian.liang@uky.edu}, Mingyang Sun$^{1}$, Yi-Bo Yang$^{2}$, Terrence Draper$^{1}$ and
Keh-Fei Liu$^{1}$\footnote[2]{liu@pa.uky.edu}}
\affiliation{$^{1}$\mbox{Department of Physics and Astronomy, University of Kentucky, Lexington, KY 40506, USA}\footnote[a]{jian.liang@uky.edu} 
$^{2}$\mbox{CAS Key Laboratory of Theoretical Physics, Institute of Theoretical Physics, Chinese Academy of Sciences, Beijing 100190, China}
%\\~\\
%\includegraphics[scale=0.2]{chiqcd.png}
}
\collaboration{$\chi$QCD Collaboration}
\begin{abstract}
The ratio of the strange quark momentum fraction $\langle x\rangle_{s+\bar{s}}$
to that of light quark $u$ or $d$ in disconnected insertions (DI) is calculated on the lattice with overlap fermions
on four domain wall fermion ensembles. These ensembles cover three lattice spacings, three volumes and several pion masses including the physical one,
from which a global fitting is carried out.
A complete nonperturbative renormalization and the mixing between the quark and glue operators
are taken into account. We find the ratio to be $\langle x\rangle_{s+\bar{s}}/\langle x\rangle_{u+\bar{u}} ({\rm DI})=0.795(79)(77)$
at $\mu = 2$ GeV in the $\overline{{\rm MS}}$ scheme. 
This ratio 
can be used as a constraint to better determine the strange parton distribution especially in the small $x$ region in the global fittings of PDFs when the connected and disconnected
sea are fitted and evolved separately, demonstrating a new way that connects lattice calculations with global analyses.
%We also compare this momentum fraction ratio with several recent global analyses of the PDF ratio  
%$(s(x) + \bar{s}(x))/(\bar{u}(x) + \bar{d}(x))$ at the same $\mu$ and discuss its consequences.
\end{abstract}

\maketitle

\section{Introduction} 
Understanding the structure of the nucleon in
terms of quarks and gluons from QCD is one of the most challenging aspects of modern nuclear and particle
physics~\cite{Geesaman:2015fha} and is of great importance in learning about how the visible Universe is built. 
Parton distribution functions (PDFs), which describe the number density of a parton with
a certain longitudinal momentum fraction $x$ and at a particular
energy scale $Q^{2}$ inside a nucleon, reveal a lot of pertinent and essential information
about the nucleon structure. In general, PDFs are determined by
global analyses of deep inelastic scattering (DIS) and Drell-Yan experiments under
the framework of QCD factorization theorems. 

For the extensively studied unpolarized PDFs, recent attention is focused on the less-known flavor structure, which
is believed to implicate the nonperturbative nature of the parton distributions due to confinement. A typical example 
%is to
%understand the origin of the Gottfried sum rule violation~\cite{Amaudruz:1991at,Towell:2001nh} which reveals that $\bar{u}(x) \neq \bar{d}(x)$.
%Another 
is the strange parton distribution which is the most uncertain among the unpolarized PDFs.
Three recent global fittings~\cite{Ball:2017nwa,Dulat:2015mca,Harland-Lang:2014zoa} with NNLO analysis show that $x (s(x)+\bar{s}(x))$ has large errors, $\sim 50\%$ or more
at $x = 10^{-3}$, and the central values of the three fits differ by $\sim 30\%$, at $Q^2 = 4\, {\rm GeV}^2$.  
%%In order to gain more precise information about the parton distributions, especially for the sea partons and gluons, in the small $x$ region,
%%proposals have been made to construct new electron-ion colliders (EIC),
%%such as the EIC in the US~\cite{Accardi:2012qut} which focuses on the glue component and the EIC China (EicC)~\cite{Chen:2019equ}
%%which plans to provide the best window to study sea quarks. Hopefully, these experiments will lead to a great leap in our understanding of the composition of a nucleon.
On the other hand, as a nonperturbative approach of solving QCD from first principles, lattice QCD could also play an important role in the study of nucleon structure.
Although there exist several pioneering approaches aiming to directly calculate the $x$ dependent PDFs on the lattice (e.g., \citep{Liu:1993cv,Liu:1998um,Liu:1999ak,Ji:2013dva,Radyushkin:2017cyf,Ma:2017pxb}), 
it is maturer and more straightforward to calculate the moments of PDFs on the lattice, which provides constraints to the PDFs.

Recent lattice calculations can already determine several quantities, e.g.\ the strange quark magnetic moment~\cite{Sufian:2016pex} and strange quark spin contribution~\cite{Liang:2018pis},
to a higher accuracy than experiments have done to date, but difficulties still exist in constraining the unpolarized strange parton distribution.
The direct difficulty is that the lattice signals of the strange quark momentum fraction $\langle x \rangle_{s+\bar{s}}$ that involves only the disconnected insertions (DIs) are not good enough \citep{Yang:2018nqn} to provide strong constraint to the global fittings.
On the other hand, lattice ratios of correlated quantities like
\begin{equation}
\label{Eq:R}
{\cal R} \equiv  \frac{\langle x\rangle_{s+\bar{s}}}{\langle x\rangle_{u+\bar{u}}({\rm DI})},
\end{equation}
where $\langle x\rangle_{u+\bar{u}}({\rm DI})$ stands for the light quark momentum fraction in disconnected insertions only (using $d$ quark makes no difference based on present lattice setup)
usually have much smaller statistical uncertainty compared with the momentum fractions themselves due to the cancellation of the statistical fluctuations of the numerator and the denominator.
%However, terminology confusion has occurred when connecting the ratio $\cal R$ to the current global fittings.
%The DI component on the lattice side \color{red}
%is part of 
%\color{black}
%the
%sea partons of global fittings in the light quark sector. The connected insertion (CI) part contains sea parton contributions as well.
However, the ratio ${\cal R}$ cannot be directly connected to the current global fittings without further theoretical insight, 
as the DI component on the lattice is part of the $u/d$ sea partons in global fittings.
%It is natural to have 
%separated connected insertions (CIs) and DIs for $u$ and $d$ quarks when 3-point correlation functions are used to calculate the moments on the lattice.
%While on the global analysis side, people are using the terms valence and sea. 
%%To tackle the difficulties, by noticing the topologically distinct path-integral diagrams of 4-point correlation functions of the hadronic tensor,
%%we find that a new classification of parton degrees of freedom~\cite{Liu:1993cv,Liu:1998um,Liu:1999ak} which reveals the sea parton contribution in the connected insertions
%%should be adopted in both the lattice community and the global analysis community. 

The parton degrees of freedom have been rigorously defined in the path-integral formulation of the hadronic tensor and classified 
according to the topologically distinct connected insertions (CIs) and DIs~\cite{Liu:1993cv,Liu:1998um,Liu:1999ak}.
This precise definition is natural for the lattice community and it is advocated to be adopted and accommodated in global fittings~\cite{Liu:2012ch, Liu:2017lpe}.
%Thereafter, lattice calculations can be directly used in and compared with global
%analyses.
Upon this basis, 
the ratio $\cal R$ has clear physical meaning and can be used as a strong 
constraint to better determine the strange quark distributions in future global fittings employing 
the path-integral
classification.
In this manuscript, we report a complete lattice
calculation of the ratio $\cal R$ at 
three lattice spacings and 
several pion masses including the physical one. Nonperturbative renormalization and the mixing from the glue momentum fraction are considered.
%\color{red}
%This can be used as a constraint in the future global fittings of PDFs to separate the CS and DS parton degrees of freedom.
%Also, more precise determination of the strange parton distribution can be made by directly using this ratio to constrain the PDF ratio of 
%$(s(x)+\bar{s}(x))/(\bar{u}(x) + \bar{d}(x))$ in the small $x$ region,
%up to the systematic uncertainty coming from the similar ratios of higher moments of the corresponding PDFs.
%\color{black}

\section{Theoretical background}
\begin{figure}[H]
\begin{centering}
\subfloat[\label{fig:1a}]{\begin{centering}
\includegraphics[scale=0.6]{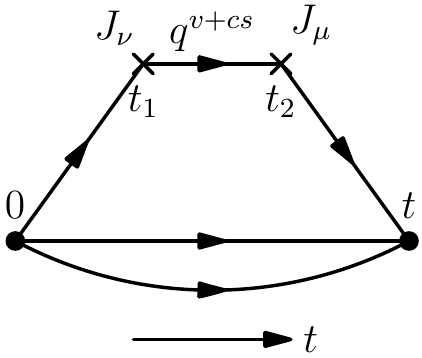}
\par\end{centering}
}\subfloat[\label{fig:1b}]{\begin{centering}
\includegraphics[scale=0.6]{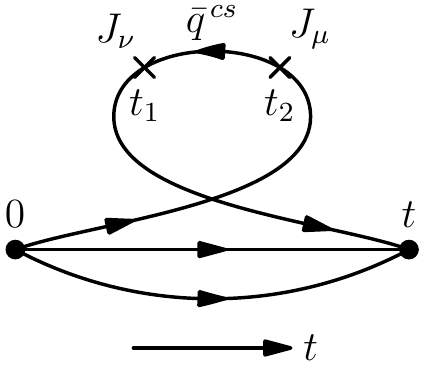}
\par\end{centering}
}\subfloat[\label{fig:1c}]{\begin{centering}
\includegraphics[scale=0.6]{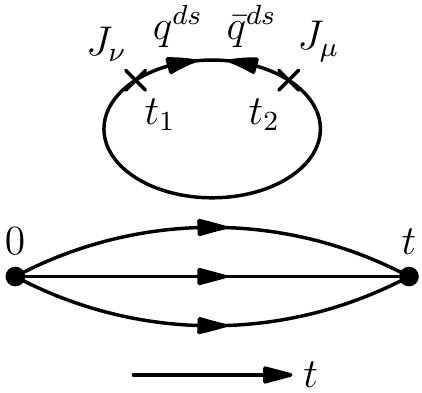}
\par\end{centering}
}
\par\end{centering}
\caption{Three topologically distinct diagrams in the Euclidean path-integral
formalism of the nucleon hadronic tensor. \label{fig:Three-topologically-distinct}}
\end{figure}

The parton
classification is revealed in the path-integral formulation of the
%%Euclidean hadronic tensor $\tilde{W}_{\mu\nu}(\vec{q},\vec{p},\tau)$~\cite{Liu:1993cv,Liu:1998um,Liu:1999ak}. 
 %
%\begin{equation}
%\tilde{W}_{\mu\nu}(\vec{q},\vec{p},\tau)\equiv\frac{E_{N}}{M_{N}}\langle p|\int\frac{d^{3}x}{2\pi}e^{-i\vec{q}\cdot\vec{x}}J_{\mu}(\vec{x},t_{2})J_{\nu}(\vec{0},t_{1})|p\rangle
%\tilde{W}_{\mu\nu}(\vec{q},\vec{p},\tau)\equiv\langle p|\int\frac{d^{3}x}{2\pi}e^{-i\vec{q}\cdot\vec{x}}J_{\mu}(\vec{x},t_{2})J_{\nu}(\vec{0},t_{1})|p\rangle
%\end{equation}
%
%%where $|p\rangle$ is the nucleon state with momentum $p$ %, $E_{N}$ and $m_{N}$ are the energy and mass of the nucleon,  
%%and $\tau=t_{2}-t_{1}$ is the time difference between the two currents $J_{\mu}$ and $J_{\nu}$. 
Euclidean hadronic tensor  $\langle p|\int\frac{d^{3}\vec{x}}{2\pi}e^{-i\vec{q}\cdot\vec{x}}J_{\mu}(\vec{x},t_{2})J_{\nu}(\vec{0},t_{1})|p\rangle$~\cite{Liu:1993cv,Liu:1998um,Liu:1999ak},
where $|p\rangle$ is a nucleon state, % , $E_{N}$ and $m_{N}$ are the energy and mass of the nucleon,  
%%and $\tau=t_{2}-t_{1}$ is the time difference between the two currents $J_{\mu}$ and $J_{\nu}$
$J_{\mu}$ and $J_{\nu}$ are two currents inserted at $t_2$ and $t_1$ and $\vec{q}$ is the momentum transfer.
Three gauge invariant and topologically distinct path-integral diagrams of the 4-point functions of the Euclidean hadronic tensor, 
which entail leading twist contributions, are illustrated in Fig.~\ref{fig:Three-topologically-distinct}.
The solid lines represent quark propagators.
The Minkowski hadronic tensor is the inverse Laplace transform of its Euclidean counterpart
and is used to extract PDFs experimentally,
%Euclidean hadronic tensor $\tilde{W}_{\mu\nu}(\vec{p},\vec{q},\tau)$~\cite{Liu:1993cv,Liu:1998um,Liu:1999ak}, where $\vec{p}$ is the
%nucleon momentum, $\vec{q}$ the momentum transfer, and $\tau$ the time difference between the two current operators. 
%The hadronic tensor $W_{\mu\nu}$ in Minkowski space
%is the inverse Laplace transform of $\tilde{W}_{\mu\nu}(\vec{q},\vec{p},\tau)$,
%i.e., 
%$W_{\mu\nu}(\vec{q},\vec{p},\nu) = 1/i \int_{c-i\infty}^{c+i\infty} e^{\nu\tau} \tilde{W}_{\mu\nu}(\vec{q},\vec{p},\tau)d\tau$.
%Three gauge invariant and topologically distinct classes of path-integral diagrams for the 4-point functions of the hadronic tensor, 
%which entail leading twist contributions, are illustrated in Fig.~\ref{fig:Three-topologically-distinct}.
%The solid lines represent quark propagators.
%%with respect to Euclidean time whereas the gluons are 
%treated as background field and 
%%not shown in the figure.
thus these diagrams of
insertions (Fig.~\ref{fig:1a} and \ref{fig:1b} are CIs and \ref{fig:1c} is DI)
classify parton degrees of freedom. %% as the separation of CI and DI in the 3-point function case.
A complete naming scheme is listed in Table~\ref{naming}.
Following this scheme, we denote Fig.~\ref{fig:1a} as $q^{v+cs}$ since in addition to the obvious valence contribution, 
there is also the connected-sea (CS) contribution coming from the higher Fock-state components 
in the Z-graph of quark lines between the two currents. %% due to the present of glue vertices (not shown in the figure). 
%Since it is in the connected insertion and produced by gluons with the quark propagating forward between the currents, we denote it as the connected-sea (CS). 
Similarly, we also have the CS antipartons ($\bar{q}^{cs}$) in Fig.~\ref{fig:1b} and the disconnected-sea (DS) partons and antipartons (${q}^{ds}+\bar{q}^{ds}$) in Fig.~\ref{fig:1c}.
%and Fig.~\ref{fig:1c} includes both the DS partons and antipartons (${q}^{ds}+\bar{q}^{ds}$).
%namely the valence and connected-sea (CS) partons ($q^{v+cs}$, Fig.~\ref{fig:1a}), 
%the CS antipartons ($\bar{q}^{cs}$, Fig.~\ref{fig:1b}), and the disconnected-sea (DS) partons and antipartons (${q}^{ds}+\bar{q}^{ds}$,  Fig.~\ref{fig:1c}).
%where the term ``sea" indicates the quarks and antiquarks created in pairs by the gluons. 
%

\begin{table}[H]
\begin{center} 
\caption{Naming scheme in the path-integral formulation of the
Euclidean hadronic tensor.\label{naming}}
\begin{tabular}{c|c}
``connected'' &  connected insertions of the Green's function\\
\hline
``disconnected'' &  disconnected insertions\\
\hline
``parton'' & forward propagating quark between currents\\
\hline
``antiparton'' & backward propagating quark\\
\hline
``valence'' & quarks from the interpolating field operators\\
\hline
``sea'' & quarks and antiquarks from gluons
\end{tabular}
\end{center}
\label{default}
\end{table}%

This is a general classification of partons
based on the continuum path-integral formulation of QCD, which is applicable to the lattice.
The nomenclature of CS and DS follows those in time-ordered perturbation theory~\cite{sterman1993introduction,schweber2011introduction}.
These two sources of sea quarks have interesting flavor dependence.
While $u$ and $d$ have both the CS
and the DS, $s$ and $c$ have only the DS.
%This is a general classification of partons
%based on the continuum path-integral formulation of QCD, which is applicable to the lattice.
%% rather than merely a lattice convention,
%%which is based on the path-integral form of QCD expressed in Euclidean space.
%One advantage of this over the canonical
%formalism is that the parton degrees of freedom are tied to the topology
%of the quark skeleton diagrams in Fig.~\ref{fig:Three-topologically-distinct},
%so that the CS and the DS can be separated.
Several experimental results demonstrate the necessity of this classification, e.g.\
%in the limit of symmetric isospin $\bar{u}^{ds}=\bar{d}^{ds}$; the
%Gottfried sum rule violation which indicates that $\bar{u}\neq\bar{d}$ comes exclusively from Fig.~\ref{fig:1b} as pointed out in~\cite{Liu:1993cv}.
%Moreover, it is 
%learned from the neutrino DIS that the momentum fraction of the strange quark is about half of that of $\bar{u}$ and $\bar{d}$, i.e.,
%$\langle x\rangle_{s+\bar{s}}/\langle x\rangle_{\bar{u}+\bar{d}}\sim0.5$, which has been incorporated in the recent global analyses~\cite{Ball:2017nwa,Dulat:2015mca,Harland-Lang:2014zoa}.
%Three recent global fittings give (the values are calculated using LHAPDF~\cite{Buckley:2014ana}) 0.56(26)~\cite{Ball:2017nwa}, 0.54(17)~\cite{Dulat:2015mca} and 0.51(16)~\cite{Harland-Lang:2014zoa}, respectively.
%On the other hand, the strange-to-down sea quark ratio $(s(x)+ \bar{s}(x))/2\bar{d}(x)$ has been determined to be around $1$ at small $x$ \cite{Aad:2012sb,Aad:2014xca}.
%This apparent dilemma can be understood in terms of the 
%fact that the DS contribution is more singular than that of the CS at small $x$~\cite{Liu:2012ch}.
the Gottfried sum rule violation is explained by the existence of CS~\cite{Liu:1993cv}.
%strange parton involves only the DS, while the $\bar{u}$ and $\bar{d}$ partons have, in addition, the CS components whose distributions are less singular than that of the disconnected sea at small $x$~\cite{Liu:2012ch}.
%measurement of the associated $W+c$ production~\cite{Aad:2014xca}.
%from an ATLAS analysis of 
%inclusive $W$ and $Z$ boson production in $pp$ collisions at LHC to be $1.00^{+0.25}_{-0.28}$ at $x$ = 0.023 and $Q^2 = 1.9\,  {\rm GeV}^2$~\cite{Aad:2012sb},
%and $0.96^{+0.26}_{-0.30}$ from the measurement of the associated $W+c$ production~\cite{Aad:2014xca}. 
%It is suggested in Ref.~\cite{Liu:2012ch} that this apparent dilemma between the ratio of momentum fractions ${\cal R}_s$ and the ratio ${r}_s(x)$ at small $x$ can be understood in terms of the 
%fact that the strange parton involves only the disconnected sea, while the $\bar{u}$ and $\bar{d}$ partons have, in addition, connected sea components
%whose distributions are less singular than that of the disconnected sea at small $x$. 
Also, under this
classification, the net valence contribution is defined as $q^v \equiv q^{v + cs}-\bar{q}^{cs}$ which is not the same as the usual definition of $q^v\equiv q-\bar{q}=q^{v + cs}+q^{ds}-\bar{q}^{cs}-\bar{q}^{ds}$ if one does not assume ${q}^{ds}=\bar{q}^{ds}$. 
%Our definition helps to avoid some ambiguities when the NNLO evolution equations are involved \cite{Liu:1999ak}.
Actually,
this definition of $q^v$ from QCD path-integral avoids some ambiguities like having a ``valence'' strange quark distribution 
when the NNLO evolution equations are involved which makes $s(x)\neq \bar{s}(x)$ \cite{Liu:1999ak}.

\begin{figure}[H]
\begin{centering}
\subfloat[\label{CI}]{\begin{centering}
\includegraphics[scale=0.6]{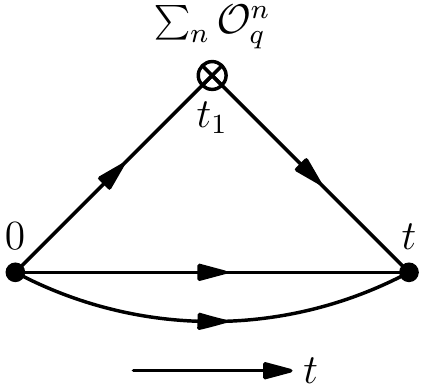}
\par\end{centering}
\hspace*{2cm}
}\subfloat[\label{DI}]{\begin{centering}
\includegraphics[scale=0.6]{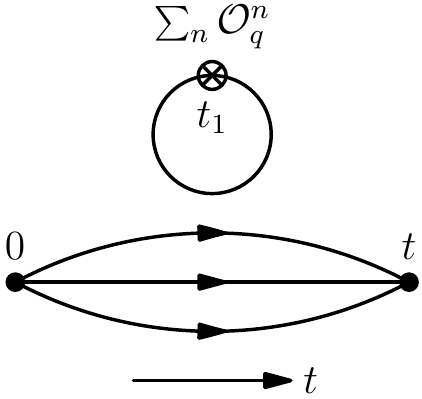}
\par\end{centering}
}
\par\end{centering}
\caption{The 3-point functions after the short-distance expansion of
the hadronic tensor from Fig.~\ref{fig:Three-topologically-distinct}. CI (a) is
derived from Fig.~\ref{fig:1a} and Fig.~\ref{fig:1b}. DI (b) originates from Fig.~\ref{fig:1c}.}
\label{fig:CIDI}
\end{figure}
Before we can study all these parton components explicitly on the lattice by directly calculating the hadronic tensor,
this classification extends the way of using common lattice calculation of 3-point functions to study the CS and DS.
It is shown~\cite{Liu:1999ak} that, upon short distance expansion, Fig.~\ref{fig:1a} together with Fig.~\ref{fig:1b} becomes
the CIs of 3-point functions in Fig.~\ref{CI} for a series of local operators $\sum_n O_q^n$,
from which the CI moments of PDFs are obtained. 
By the same token, the disconnected
4-point functions in Fig.~\ref{fig:1c} become the DIs of 3-point functions
in Fig.~\ref{DI} to obtain the DI moments.
The DI ratio $\cal R$ therefore represents the DS ratio of the strange quark to light quark, containing the information
needed in global fittings to separate CS and DS and improve the strange parton results.

\begin{table} [H]
\centering{}\caption{Parameters of the RBC/UKQCD ensembles: spatial/temporal
size, lattice spacing, sea strange quark mass under $\overline{\textrm{MS}}$
scheme at 2 GeV, pion mass with degenerate light sea quark, and
the number of configurations. \label{tab:The-parameters}}
\begin{tabular}{cccccc}
Symbol & $L^{3}\times T$ & $a$(fm)  & $m_{s}^{s}$(MeV) & $m_{\pi}^{s}$(MeV)  & $N_{{\rm cfg}}$\tabularnewline
\hline 
\hline 
32ID & $32^{3}\times64$ & 0.1431(7) & 89.4  & 171 & 200\tabularnewline
\hline 
24I & $24^{3}\times64$ & 0.1105(3) & 120 & 330 & 203\tabularnewline
\hline 
48I & $48^{3}\times96$ & 0.1141(2) & 94.9 & 139 & 81\tabularnewline
\hline 
32I & $32^{3}\times64$ &  0.0828(3) & 110 & 300 & 309\tabularnewline
\hline 
\end{tabular}
\end{table}

\section{Numerical Details}

\begin{table*}[htbp]
\caption{The details of the overlap simulation in the valence sector for the
CI case, including the name of the lattice, the grid type of source
$\mathcal{G}_{{\rm src}}$ (the notations such as 12-12-12 denote
the intervals of the grid in the three spatial directions; see
reference \citep{Yang:2015zja} for more details), the number of noises for the source
grids $N_{{\rm src}}$, the time positions of sources $t_{{\rm src}}$,
the grid type of sink $\mathcal{G}_{{\rm sink}}$, the number of noises for the sink
grids $N_{{\rm sink}}$, the source-sink separations $(t_{{\rm sink}}-t_{{\rm src}})$,
and the bare valence quark masses $m_{q}^{v}a$. \label{tb:valence_setup_CI}}
\centering{}%
\begin{tabular}{r@{\extracolsep{0pt}.}lr@{\extracolsep{0pt}.}lr@{\extracolsep{0pt}.}lr@{\extracolsep{0pt}.}lr@{\extracolsep{0pt}.}lccc}
\hline 
\multicolumn{2}{c}{Lattice} & \multicolumn{2}{c}{$\mathcal{G}_{{\rm src}}$} & \multicolumn{2}{c}{$N_{{\rm src}}$} & \multicolumn{2}{c}{$t_{{\rm src}}$} & \multicolumn{2}{c}{$\ensuremath{\mathcal{G}_{{\rm sink}}}$} & $N_{{\rm sink}}$ & $(t_{{\rm sink}}-t_{{\rm src}})$ & $m_{q}^{v}a$\tabularnewline
\hline 
\multicolumn{2}{c}{} & \multicolumn{2}{c}{} & \multicolumn{2}{c}{} & \multicolumn{2}{c}{} & \multicolumn{2}{c}{} & 5 & 0.88 fm & \tabularnewline
\cline{11-12} 
\multicolumn{2}{c}{24I} & \multicolumn{2}{c}{12-12-12} & \multicolumn{2}{c}{1} & \multicolumn{2}{c}{(0, 32)} & \multicolumn{2}{c}{2-2-2} & 5 & 1.11 fm & (0.0102, 0.0135, 0.0160, 0.0203) \tabularnewline
\cline{11-12} 
\multicolumn{2}{c}{} & \multicolumn{2}{c}{} & \multicolumn{2}{c}{} & \multicolumn{2}{c}{} & \multicolumn{2}{c}{} & 5 & 1.33 fm & \tabularnewline
\hline 
\multicolumn{2}{c}{} & \multicolumn{2}{c}{} & \multicolumn{2}{c}{} & \multicolumn{2}{c}{} & \multicolumn{2}{c}{} & 3 & 0.99 fm & \tabularnewline
\cline{11-12} 
\multicolumn{2}{c}{32I} & \multicolumn{2}{c}{16-16-16} & \multicolumn{2}{c}{1} & \multicolumn{2}{c}{(0, 32)} & \multicolumn{2}{c}{1-1-1} & 3 & 1.16 fm & (0.00765, 0.00885, 0.0112, 0.0152)\tabularnewline
\cline{11-12} 
\multicolumn{2}{c}{} & \multicolumn{2}{c}{} & \multicolumn{2}{c}{} & \multicolumn{2}{c}{} & \multicolumn{2}{c}{} & 3 & 1.24 fm & \tabularnewline
\hline

\multicolumn{2}{c}{} & \multicolumn{2}{c}{} & \multicolumn{2}{c}{} & \multicolumn{2}{c}{} & \multicolumn{2}{c}{} & 4 & 1.29 fm & \tabularnewline
\cline{11-12} 
\multicolumn{2}{c}{32ID} & \multicolumn{2}{c}{16-16-16} & \multicolumn{2}{c}{6} & \multicolumn{2}{c}{(0, 32)} & \multicolumn{2}{c}{1-1-1} & 5 & 1.43 fm & \footnotesize{(0.0042, 0.0060, 0.011, 0.014, 0.017, 0.022)} \tabularnewline
\cline{11-12} 
\multicolumn{2}{c}{} & \multicolumn{2}{c}{} & \multicolumn{2}{c}{} & \multicolumn{2}{c}{} & \multicolumn{2}{c}{} & 12 & 1.57 fm & \tabularnewline
\hline 
\multicolumn{2}{c}{} & \multicolumn{2}{c}{} & \multicolumn{2}{c}{} & \multicolumn{2}{c}{} & \multicolumn{2}{c}{} & 4 & 0.88 fm & \tabularnewline
\cline{11-12} 
\multicolumn{2}{c}{48I} & \multicolumn{2}{c}{12-12-12} & \multicolumn{2}{c}{5} & \multicolumn{2}{c}{(0, 32, 64)} & \multicolumn{2}{c}{1-1-1} & 8 & 1.11 fm & \scriptsize{(0.0024, 0.0030, 0.00809, 0.0102, 0.0135, 0.0160, 0.0203)} \tabularnewline
\cline{11-12} 
\multicolumn{2}{c}{} & \multicolumn{2}{c}{} & \multicolumn{2}{c}{} & \multicolumn{2}{c}{} & \multicolumn{2}{c}{} & 12 & 1.33 fm & \tabularnewline
\hline 
\end{tabular}
\end{table*}
%The numerical setup of this study is
%the same as in our previous work \citep{Yang:2018nqn}. 
We use overlap
fermions \citep{Neuberger:1997fp} as valence quarks on four $2+1$-flavor
RBC/UKQCD gauge ensembles with domain wall fermions \citep{Aoki:2010yq,Blum:2014tka}.
The parameters of the ensembles are listed in Table \ref{tab:The-parameters}.
We have three different lattice spacings and lattice volumes respectively,
and four values of sea pion mass with one at the physical point.
For the valence sector, multiple partially-quenched valence quark
masses are used, owing to the multi-mass algorithm. We choose four
valence quark masses ranging from $\sim250$ to $\sim400$ MeV on
the 24I and 32I ensembles and 7/6 quark masses in the range $\left[130, 400\right]$
MeV on the 48I/32ID ensemble. Combining these ensembles and valence pion masses
in a global analysis helps to control the lattice systematic uncertainties
and leads to our final result at the physical limit.

The quark and glue momentum fractions in the nucleon can be defined
by the matrix element of the traceless diagonal part of the energy-momentum
tensor (EMT) in the rest frame \citep{Horsley:2012pz},
\begin{equation}
\langle x\rangle_{q,g}\equiv-\frac{\langle N|\frac{4}{3}\overline{T}_{44}^{q,g}|N\rangle}{M_{N}\langle N|N\rangle},
\end{equation}
with $\overline{T}_{44}^{q}=\int d^{3}x\overline{\psi}(x)\frac{1}{2}\left(\gamma_{4}\overleftrightarrow{D}_{4}-\frac{1}{4}{\displaystyle \sum_{i=0,1,2,3}}\gamma_{i}\overleftrightarrow{D}_{i}\right)\hat{\psi}(x)$
and $\overline{T}_{44}^{g}=\int d^{3}x\frac{1}{2}\left[E(x)^{2}-B(x)^{2}\right].$
Here $\hat{\psi}=(1-\frac{1}{2}D_{{\rm ov}})\psi$ is for giving rise
to the effective quark propagator $\left(D_{c}+m\right)^{-1}$,
where $D_{c}$ satisfying $\mbox{\ensuremath{\left\{ D_{c},\gamma_{5}\right\} =0}}$
is exactly chiral and can be defined from the original overlap operator
$D_{{\rm ov}}$ as $D_{c}=\frac{\rho D_{{\rm ov}}}{1-D_{{\rm ov}}/2}$ \citep{Liu:2002qu}.
More details regarding the calculation of the overlap operator and
eigenmodes deflation in the inversion of the fermion matrix can be
found in \citep{Li:2010pw}. To calculate the matrix elements, we
need first to construct 3-point correlation functions 
\begin{equation}
C_{3}^{q,g}(t_{f},\tau)=\sum_{\vec{x},\vec{y}}\langle\chi(t_{f},\vec{y})\overline{T}_{44}^{q,g}(\tau,\vec{x})\bar{\chi}(0,\mathcal{G})\rangle,
\end{equation}
where $\chi$ is the nucleon interpolation field and $\mathcal{G}$ denotes
the source grid. Then, we make a ratio of the 3-point correlation
function to the nucleon 2-point function and extract the matrix element by fitting the ratio
using the so-called two-state form
\begin{equation} \label{two-state}
\begin{split}
&{\Pi}^{q,g}(t_{f},\tau)=\frac{{\rm Tr}\left[\Gamma_{e}C_{3}^{q,g}(t_{f},\tau)\right]}{{\rm Tr}\left[\Gamma_{e}C_{2}(t_{f})\right]}\\
=&\langle N|\overline{T}_{44}^{q,g}|N\rangle+c_{1}^{q,g}e^{-\delta m\left(t_{f}-\tau\right)}+c_{2}^{q,g}e^{-\delta m\tau}+c_{3}^{q,g}e^{-\delta mt_{f}}.
\end{split}
\end{equation}
%such that $\langle N|\overline{T}_{44}^{q,g}|N\rangle={\Pi}^{q,g}(t_{f}\gg\tau,\tau\gg0)$.
Here $\Gamma_{e}$ is the non-polarized projector, $C_{2}(t_{f})=\sum_{\vec{x}}\langle\chi(t_{f},\vec{x})\bar{\chi}(0,\mathcal{G})\rangle$,
%At finite $t_{f}$ and $\tau$, the excited states will also contribute
%to the matrix element and we need to extract it by fitting the ratio
%by the so-called two-state form
%\begin{equation}   \label{two-state}
%{\Pi}^{q,g}(t_{f},\tau)=c_{0}^{q,g}+c_{1}^{q,g}e^{-\delta m\left(t_{f}-\tau\right)}+c_{2}^{q,g}e^{-\delta m\tau}+c_{3}^{q,g}e^{-\delta mt_{f}},
%\end{equation}
$c$'s are fitting coefficients, and $\delta m$ is the effective energy difference between the ground
state and the excited states. To better use this formula, multiple
source-sink separations $t_{f}$ ranging from $\sim0.7$ fm to $\sim1.5$
fm are constructed for ${\Pi}^{q,g}(t_{f},\tau)$ on each ensemble for all
the current positions $\tau$ between the source and sink. 
%More detailed
%examples of two-state fits can be found in our previous works, e.g. \citep{Yang:2016plb,Liang:2018pis}.

As mentioned above, the 3-point correlation functions have two kinds
of current insertions, CI and DI, as illustrated in Fig.~\ref{fig:CIDI}. 
%So the ratios and thus the matrix elements are calculated separately as well. 
Since both the CI and glue matrix elements mix to DI through the renormalization of bare quantities 
under lattice regularization~\cite{Yang:2018nqn}, 
the calculation of the ratio ${\cal R}$ under $\overline{\rm MS}$ scheme involves also
the CI and glue contributions.

For the CI calculations,
we use the stochastic sandwich method (SSM)~\citep{Yang:2015zja} with low-mode substitution
(LMS)~\cite{Li:2010pw} to better control the statistical uncertainty.
$Z_{3}$-noise grid sources with Gaussian smearing (for the 48I, 24I and 32I lattices) or
block smearing~\citep{Liang:2016fgy} (for the 32ID lattice) are placed coherently at $t_{{\rm {src}}}=0$
and $t_{{\rm {src}}}=32$ ($t_{{\rm {src}}}=64$ also for 48I) in one inversion. 
Nucleon sinks are located
at different positions with different separations in time from the source.
Technical details regarding
the LMS of a random $Z_{3}$ grid source and the use of SSM with LMS for constructing
3-point functions can be found in Refs.~\citep{Gong:2013vja,Yang:2015zja,Liang:2016fgy}.
Setups regarding the valence sector of the CI case are listed
in Table~\ref{tb:valence_setup_CI}.
Due to the fact that the multi-mass inversion algorithm is applicable to
the overlap fermion with eigenvector deflation, 
we calculate four to seven valence masses for each of the four lattices. 
%Technical details regarding
%the LMS of random $Z_{3}$ grid source and the SSM with LMS for constructing
%3-point functions can be found in Refs.~\citep{Gong:2013vja,Yang:2015zja,Liang:2016fgy}.

For the DI calculations, we use the low-mode average
(LMA) technique to calculate the quark loops which improves the signal-to-noise
ratios.
The low-mode part of the quark loops is calculated exactly since we have solved the 
low-lying eigenvectors of the overlap Dirac operator on all these lattices,
The high-mode part is estimated
with 8 sets of $Z_{4}$-noise on a 4-4-4-2 space-time grid with even-odd dilution
and additional time shift (32 inversions in total). 
The same smeared $Z_{3}$-noise grid sources as used in the CI case 
are used in the production of the nucleon
propagators. 
We make multiple measurements by shifting the source
along the time direction to improve statistics. 
The spatial position of the
center of the grid is randomly chosen for each source at different times to
reduce autocorrelation. 
References~\citep{Gong:2013vja,Gong:2015iir,Yang:2015uis}
contain more details regarding the DI calculation. 
When constructing
quark loops, we include more valence quark masses to cover the strange
region. 
The bare valence strange quark masses are determined on each lattice
by the global-fit
value at 2 GeV in the $\overline{{\rm MS}}$ scheme calculated in our previous study~\citep{Yang:2014sea} 
and the nonperturbative mass renormalization
constants calculated in~\citep{Liu:2013yxz};
corresponding numbers are collected in Table~\ref{tb:strange_DI}.
For all the 24I, 48I and 32I lattices the renormalized strange quark mass
is around 100.5 MeV and for the 32ID lattice the number is around 95 MeV, which are all 
consistent with our global-fit
value 101(3)(6) MeV~\citep{Yang:2014sea} within error.
We used the clover definition of the glue operator~\citep{Yang:2018bft} 
for the DI calculation of the glue momentum fraction.
The cluster-decomposition error reduction (CDER) technique is applied
to improve the signal~\citep{Liu:2017man,Yang:2018bft}.

\begin{table}[H]
\caption{The bare valence strange quark mass parameters and mass renormalization constants ($\overline{\rm MS}$ at 2 GeV)
used in DI. \label{tb:strange_DI}}
\centering{}%
\begin{tabular}{ccccc}
\hline 
& 32I & 24I & 48I & 32ID\\
\hline 
$m_sa$ & 0.04454 & 0.06347 & 0.06548 & 0.08500\\
\hline 
$Z_m$ & 0.9467(57) & 0.8872(68) & 0.8872(68) & 0.8094(26) \\
\hline 
\end{tabular}
\end{table}

%References~\citep{Gong:2013vja,Gong:2015iir,Yang:2015uis}
%contain more details regarding the DI calculation. 

\section{Renormalization}

\begin{table}[H]
\begin{center}
\caption{\label{table:Z} The nonperturbative renormalization constants on different ensembles at $\overline{\textrm{MS}}$ \mbox{2~GeV}. The 24I and 48I ensembles share the same renormalization constants due to the same lattice spacing.}
\begin{tabular}{cccc}
Symbol & $Z_{QQ}^{\overline{\textrm{MS}}}$  & $\delta Z_{QQ}^{\overline{\textrm{MS}}}$ & $Z_{QG}^{\overline{\textrm{MS}}}$  \\
\hline
32ID & 1.25(0)(2) & 0.018(2)(2) & 0.017(17)  \\
24I/48I & 1.24(0)(2) & 0.012(2)(2) & 0.007(14)  \\
32I &  1.25(0)(2) & 0.008(2)(2) & 0.000(14) \\
\hline
\end{tabular}
\end{center}
\end{table}
%The comprehensive nonperturbative renormalization
%used in this work for both the quark and glue sections is formulated
%and implemented in \citep{Yang:2018bft,Yang:2018nqn}.
As demonstrated
in \citep{Liang:2018pis}, the renormalization can be processed separately
for CI and DI and we will focus on the DI part in this work. 
The general form 
of the renormalized
momentum fractions in DI $\langle x\rangle^{R,{\rm DI}}$ in the $\overline{\textrm{MS}}$
scheme at scale $\mu$ reads
\begin{align}
\langle x\rangle_{u,d,s}^{R,{\rm DI}} & =Z_{QQ}^{\overline{\textrm{MS}}}(\mu)\langle x\rangle_{u,d,s}^{{\rm DI}}+\nonumber \\
+ & \delta Z_{QQ}^{\overline{\textrm{MS}}}(\mu)\sum_{q=u,d,s}\langle x\rangle_{q}^{{\rm CI+DI}}+Z_{QG}^{\overline{\textrm{MS}}}(\mu)\langle x\rangle_{g},
\end{align}
where $\langle x\rangle_{u,d,s}^{{\rm DI/CI}}$ is the bare quark momentum
fraction in the DI/CI sector under lattice regularization, $\langle x \rangle_{g}$ is the glue momentum fraction, $Z_{QQ}^{\overline{\textrm{MS}}}(\mu)$ is the renormalization constant and 
$ \delta Z_{QQ}^{\overline{\textrm{MS}}}(\mu)$ and $Z_{QG}^{\overline{\textrm{MS}}}(\mu)$ account for the mixing. 
To renormalize a lattice-regularized quantity with ${\overline{\textrm{MS}}}$ scheme, we first use the RI/MOM scheme to renormalize
it at a scale $\mu_R$ nonperturbatively. 
And then,
we convert the RI/MOM renormalized quantity to the ${\overline{\textrm{MS}}}$ scheme using 
a perturbatively calculated matching coefficient and evolve it to certain scale $\mu$. 
The complete renormalization is a combination of these two steps and can be expressed formally as 
$Z^{\overline{{\rm MS}}}(\mu)=\left[\left( Z(\mu_{R}) R(\mu/\mu_{R}) \right) |_{a^{2}\mu_{R}^{2}\rightarrow0}\right]^{-1}$,
where $Z(\mu_{R})$ and $R(\mu/\mu_{R})$ denote the RI/MOM renormalization and matching respectively.
%Please note that one needs to carry out the RI/MOM renormalization 
%with several quark masses on each lattice
%and extrapolate the results to
%the massless limit before the matching due to the use of massless renormalization scheme.
%Example plots and discussion about systematic uncertainties can be found in Sec.~II
%of the supplemental materials~\cite{sup-disratio:2020aaa}.

\begin{figure}[h]
\subfloat[24I\label{chiral_renorm_24I}]{
\includegraphics[scale=0.48]{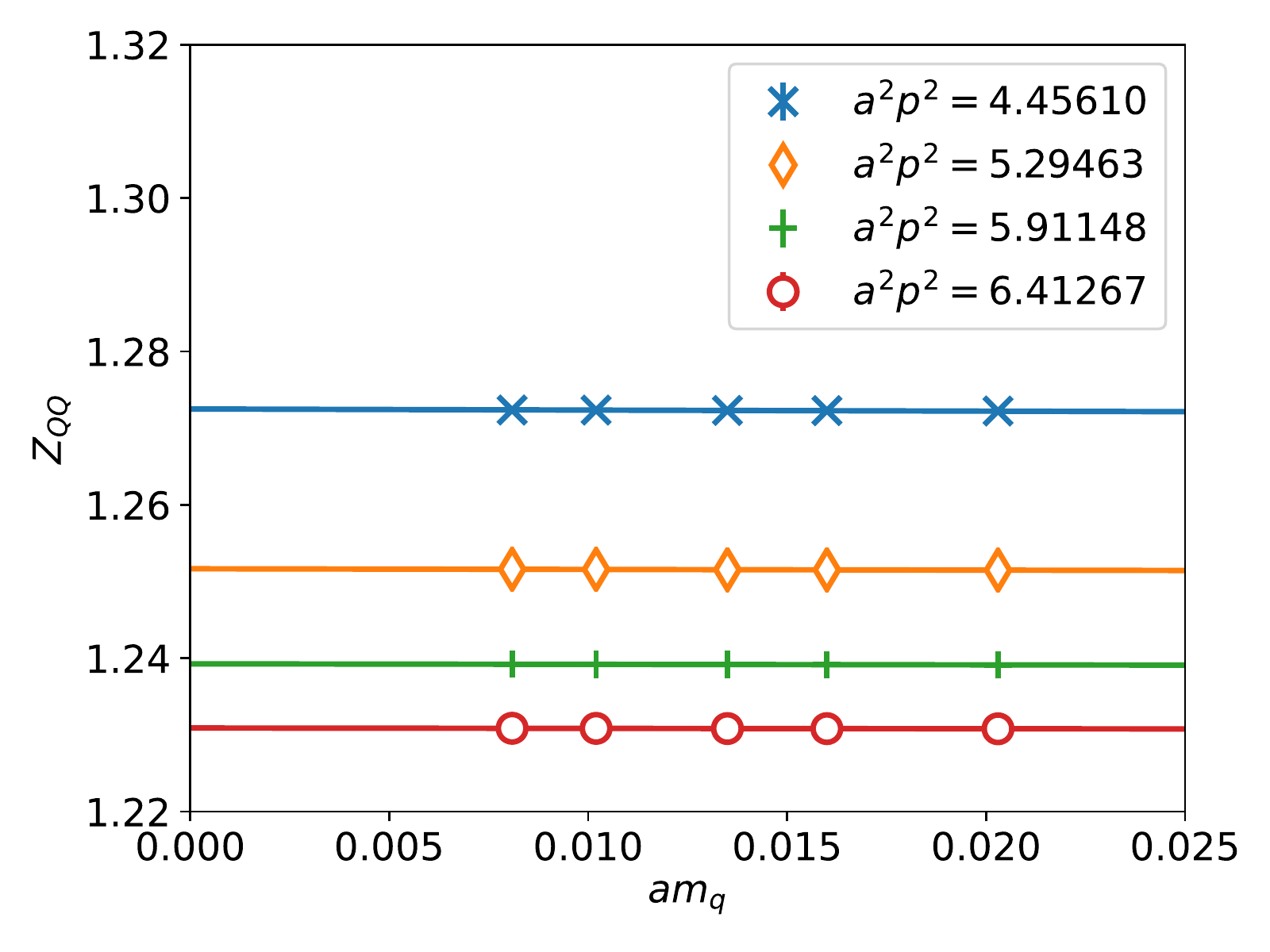}
}\\
\subfloat[32ID\label{chiral_renorm_32ID}]{
\includegraphics[scale=0.48]{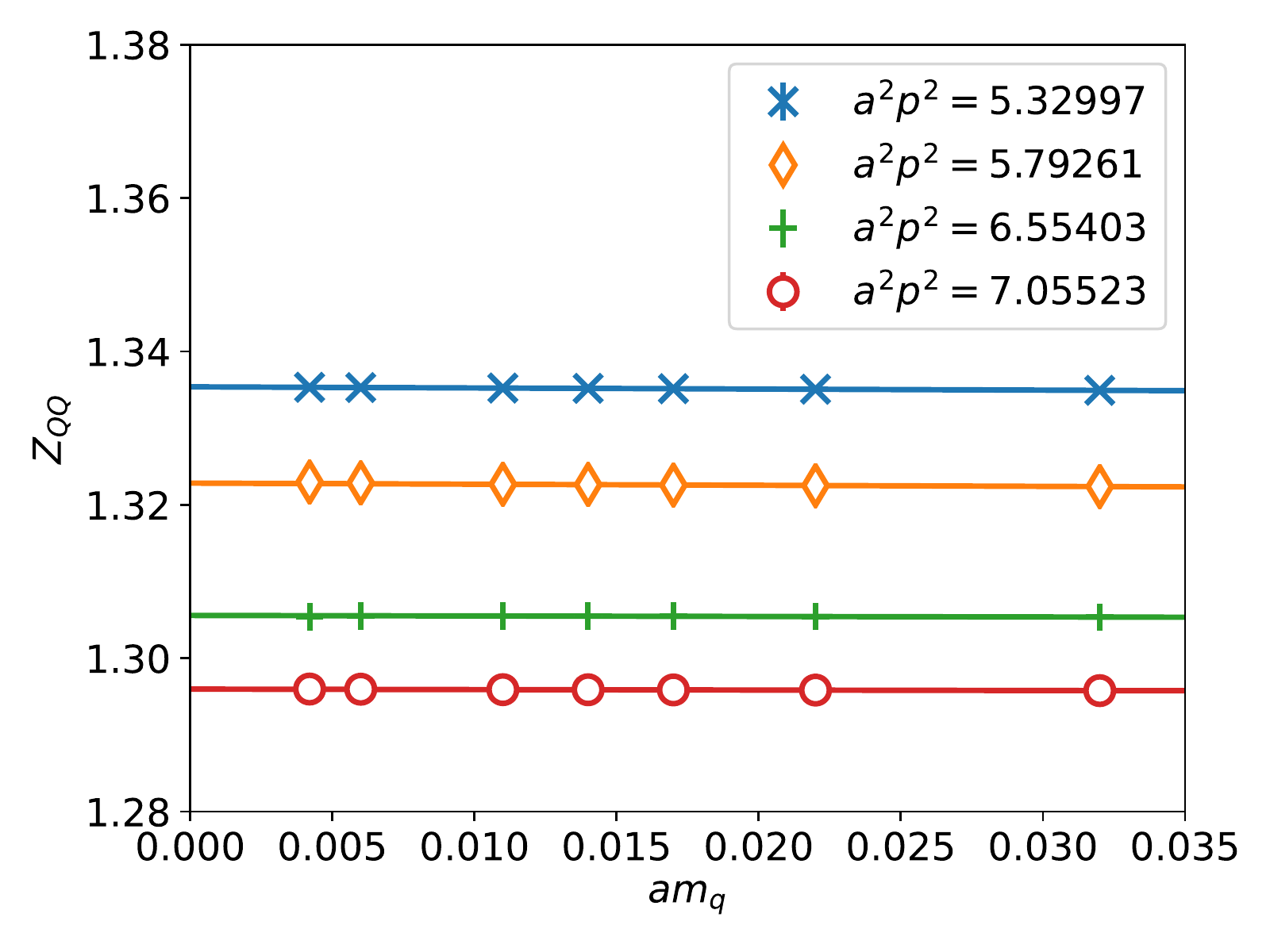}
}
\caption{The isovector RI/MOM renormalization constants of the
traceless diagonal part of the EMT at different scales $a^2p^2$ 
as a function of the bare valence quark mass.
The left panel is for the 24I lattice and the right one is
for the 32ID lattice.
Solid lines show linear extrapolations to the massless limit.\label{chiral_renorm}}
\end{figure}

In the nonperturbative renormalization procedure,
one needs to carry out the RI/MOM renormalization several times 
with several quark masses on each lattice
and extrapolate the results to
the massless limit before the matching from RI/MOM to $\overline{\rm MS}$
since the massless renormalization scheme is used.
Example plots for the
isovector RI/MOM renormalization constants of the
traceless diagonal piece of the EMT $Z_{QQ}$ as a function of the bare valence quark mass
are shown in Fig.~\ref{chiral_renorm}.
Different colors denote different $a^2p^2$ scales.
Fig.~\ref{chiral_renorm_24I} is for the 24I lattice and Fig.~\ref{chiral_renorm_32ID}
is for the 32ID lattice.
It can be observed from the figures that the
quark mass dependence is mild for both lattices at all scales,
and linear fits (the solid lines in the figure) can be used to extrapolate
the results to the massless limit.

In principle, this kind of extrapolation needs to be done also for the mixing
coefficients. However, in practice, it is not necessary since from Fig.~\ref{chiral_renorm_24I} we
learned that the finite quark mass effect is quite small compared with the large statistical errors of the 
mixing coefficients.
Also, we did not extrapolate the sea quark masses to zero.
But our previous work~\cite{Liu:2013yxz},
where we used three 24I and three 32I lattices with different sea quark masses 
and several valence quark masses on each lattice
to take both the valence and sea quark mass to zero for the massless renormalization scheme,
shows 
that the sea quark mass effects are usually very weak for other renormalization constants (e.g., $Z_A$, $Z_P$ and $Z_S$).
Accordingly, we can estimate
the systematic uncertainty of the renormalization constants of this work due to the non-zero sea masses.
%Actually, a full error budget of the renormalization constants was listed in Table~I of 
%the supplementary materials of our previous work~\cite{Yang:2018nqn},
%and all the systematic uncertainties of the renormalization constants are included in the present calculations.
Actually, our previous work~\cite{Yang:2018nqn}
has estimated all the systematic uncertainties of the renormalization constants of EMT on these lattices
and a full error budget can be found in its supplementary materials.
All those systematic uncertainties of the renormalization constants are included in the present calculations.

%To renormalize a lattice-regularized quantity with ${\overline{\textrm{MS}}}$ scheme, we usually first use the RI/MOM scheme to renormalize
%it at the scale $\mu_R$ and the corresponding renormalization constant and mixing coefficients are denoted by $Z_{QQ}(\mu_{R})$, $Z_{GG}(\mu_{R})$, $Z_{GQ}(\mu_{R})$ and $Z_{QG}(\mu_{R})$ respectively.
%Then, we convert the renormalized quantity from the RI/MOM scheme to the ${\overline{\textrm{MS}}}$ scheme and evolve it to some scale $\mu$. 
%The matching coefficients are denoted by $R_{QQ}(\frac{\mu}{\mu_R})$, $R_{GG}(\frac{\mu}{\mu_R})$, $R_{GQ}(\frac{\mu}{\mu_R})$ and $R_{QG}(\frac{\mu}{\mu_R})$. The complete renormalization is a combination of these two steps and can be %expressed in the following matrix form 
%\color{black}
%\begin{widetext}
%\begin{eqnarray}
%\left(\begin{array}{cc}
%Z_{QQ}^{\overline{\textrm{MS}}}(\mu)+N_{f}\delta Z_{QQ}^{\overline{\textrm{MS}}}(\mu) & N_{f}Z_{QG}^{\overline{\textrm{MS}}}(\mu)\\
%Z_{GQ}^{\overline{\textrm{MS}}}(\mu) & Z_{GG}^{\overline{\textrm{MS}}}(\mu)
%\end{array}\right) & \equiv & \left\{ \left[\left(\begin{array}{cc}
%Z_{QQ}(\mu_{R})+N_{f}\delta Z_{QQ} & N_{f}Z_{QG}(\mu_{R})\\
%Z_{GQ}(\mu_{R}) & Z_{GG}(\mu_{R})
%\end{array}\right)\right.\right.\nonumber \\
% &  & \left.\left.\left.\left(\begin{array}{cc}
%R_{QQ}(\frac{\mu}{\mu_{R}})+{\cal O}(N_{f}\alpha_{s}^{2}) & N_{f}R_{QG}(\frac{\mu}{\mu_{R}})\\
%R_{GQ}(\frac{\mu}{\mu_{R}}) & R_{GG}(\frac{\mu}{\mu_{R}})
%\end{array}\right)\right]\right|_{a^{2}\mu_{R}^{2}\to0}\right\} ^{-1},
%\end{eqnarray}
%\end{widetext} 
We use the 3-loop result for the isovector matching
coefficient~\citep{Gracey:2003mr}
while only 1-loop results exist for the others~\citep{Yang:2016xsb}.
More detailed discussion about the calculation of nonperturbative renormalization and mixing is beyond the scope of this paper,
and can be found in our previous works~\citep{Yang:2018bft,Yang:2018nqn}.
The renormalization constants used in this work are listed in Table~\ref{table:Z}.
%One can see that the mixing effects are not that strong with our lattice setup which are of order $0.01$ or less.
Although the mixing coefficients are of order $0.01$ or less, 
the mixing effects of this study are significant ($\sim 10\%$)
since the DI bare values are themselves smaller than those of CI and glue.

\section{Results} 
The two bare matrix elements of the strange and
light quarks 
%of each valence pion mass on each ensemble 
are fitted
using the two-state formula (Eq.~(\ref{two-state})) in a joint correlated fit, such that
the correlation between the two matrix elements is properly kept.
%Although either of the matrix elements is relatively noisy, the ratio
%has small statistical error due to the fact that the statistical fluctuations
%are reduced via correlation.
This ensures the cancellation of the fluctuations of the two matrix elements in the ratio 
and leads to statistically more stable results.

\begin{figure*}[htbp]
\subfloat[light quark\label{}]{
\includegraphics[scale=0.33]{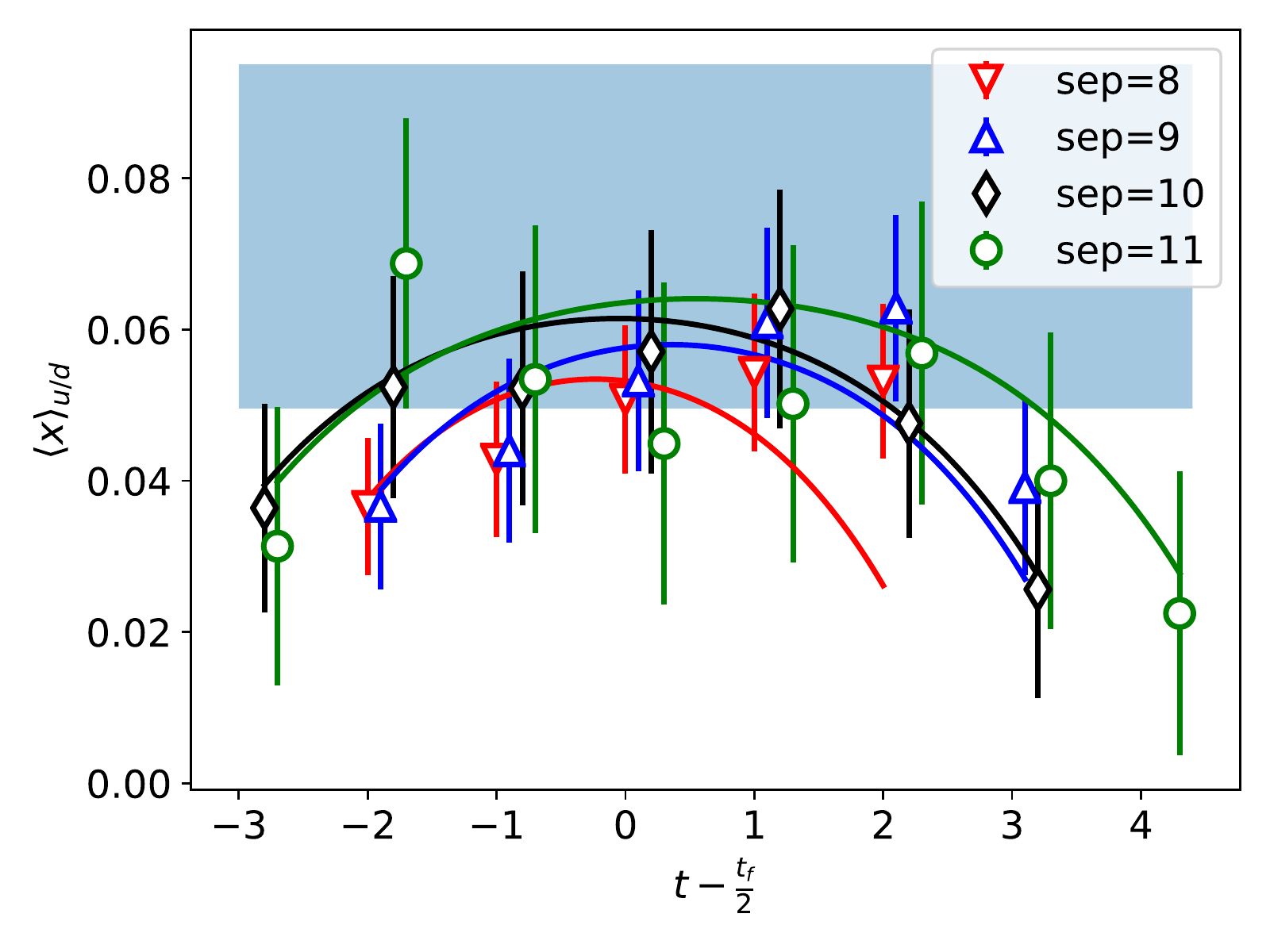}
}
\subfloat[strange quark\label{}]{
\includegraphics[scale=0.33]{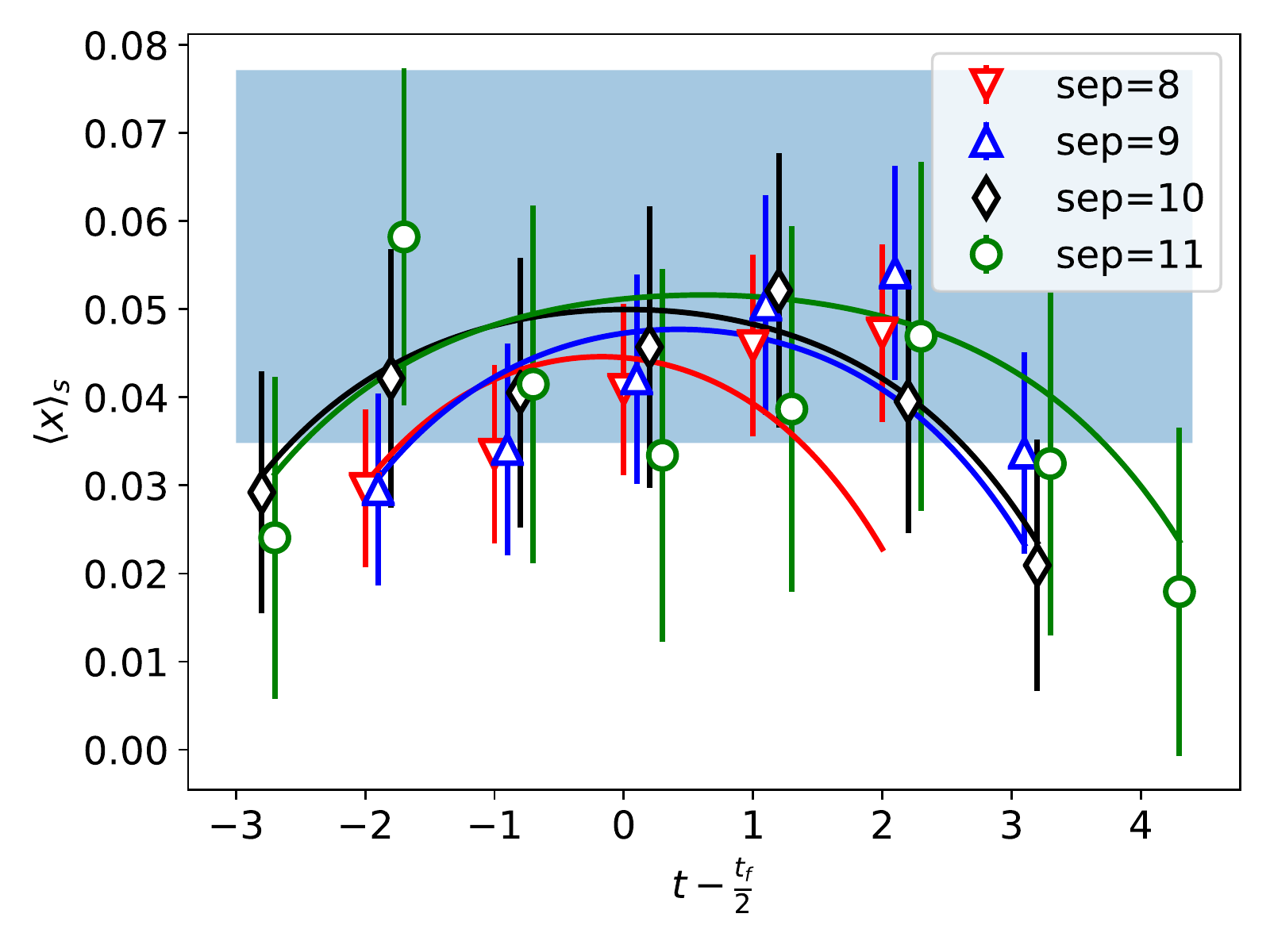}
}
\subfloat[ratio\label{}]{
\includegraphics[scale=0.33]{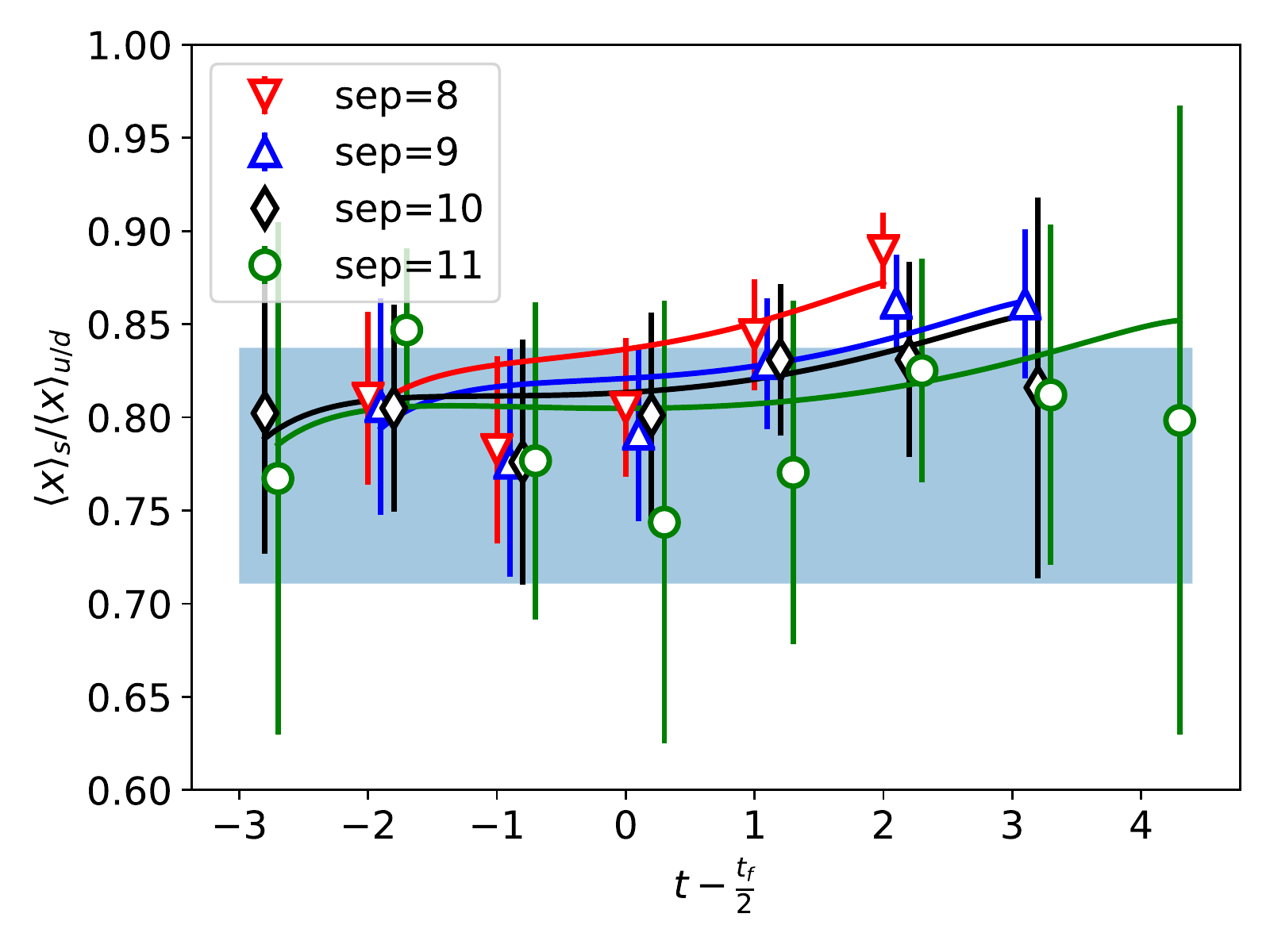}
}
\caption{Example plots of the momentum fractions in DI on the 24I lattice at its unitary point.
Panels~(a), (b) and (c) show the results of the light quark, strange quark and
the bare ratio ${\cal R}$ respectively. Different colors denote different source-sink separations.
Colored lines are the two-state fittings for each separation and the blue bands
indicate the final results at infinite source-sink separations.\label{DI1}}
\end{figure*}

Two-state fits are employed to handle 
the excited-state contaminations at finite 
source-sink separations.
Fig.~\ref{DI1} shows
example plots of the momentum fractions in DI on the 24I lattice at its unitary point.
Different colors denote different source-sink separations.
The colored lines in Panels~(a) and (b) show the two-state fittings for each separation 
and their ratios are plotted in Panel (c).
The blue bands indicate the final results at infinite source-sink separations.
We use a joint fit involving both the light and strange quark, such that
the ratio of the strange quark momentum fraction to that of light quark has
much smaller relative error than the momentum fraction themselves due to the cancellation of the 
statistical fluctuations.
All the fittings result in good $\chi^2/d.o.f.$ (around or less than one) and
the final errors represented by the height of the bands are similar to the errors of
the data points at the largest source-sink separation, which is reasonable and reassuring.
Detailed fitting setups can be found in Table~\ref{tab:2s_fitting_DI}.

\begin{table}[htbp]
\caption{Setups of the two-state fits in the DI part. The source-sink separations
used in the fits, the number of points dropped on the source side and the sink side,
and the prior value
and width of $\delta m$ are listed for each lattice.\label{tab:2s_fitting_DI}}
\begin{centering}
\begin{tabular}{ccccc}
\cline{1-5}
\multicolumn{1}{c}{lattice} &\multicolumn{1}{c}{separations (a)}&\multicolumn{1}{c}{source drop}
&\multicolumn{1}{c}{sink drop}&\multicolumn{1}{c}{prior $\delta ma$}\tabularnewline
\hline 
32ID & 6, 7, 8 & 1 & 1 & 0.4($\infty$)\tabularnewline
\hline 
24I & 8, 9, 10, 11 & 2 & 1 & 0.3($\infty$)\tabularnewline
\hline 
48I & 6, 7, 8 & 1 & 1 & 0.3(0.6)\tabularnewline
\hline 
32I & 8, 9, 10, 11 & 2 & 3 & 0.2($\infty$)\tabularnewline
\hline 
\end{tabular}
\par\end{centering}
\end{table}

\begin{figure}[htbp]
\subfloat[24I]{
\includegraphics[scale=0.48]{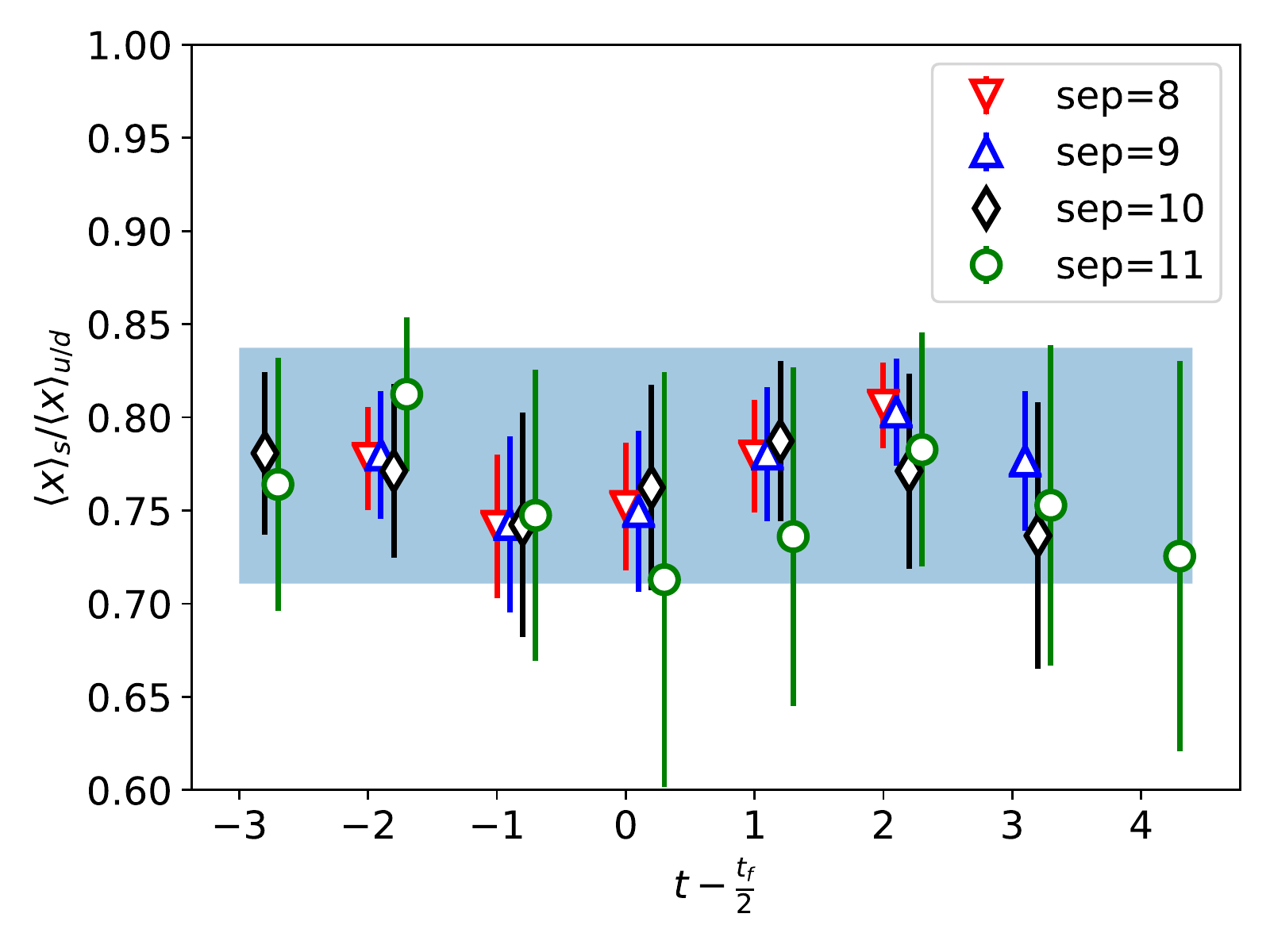}
}\\
\subfloat[32I\label{}]{
\includegraphics[scale=0.48]{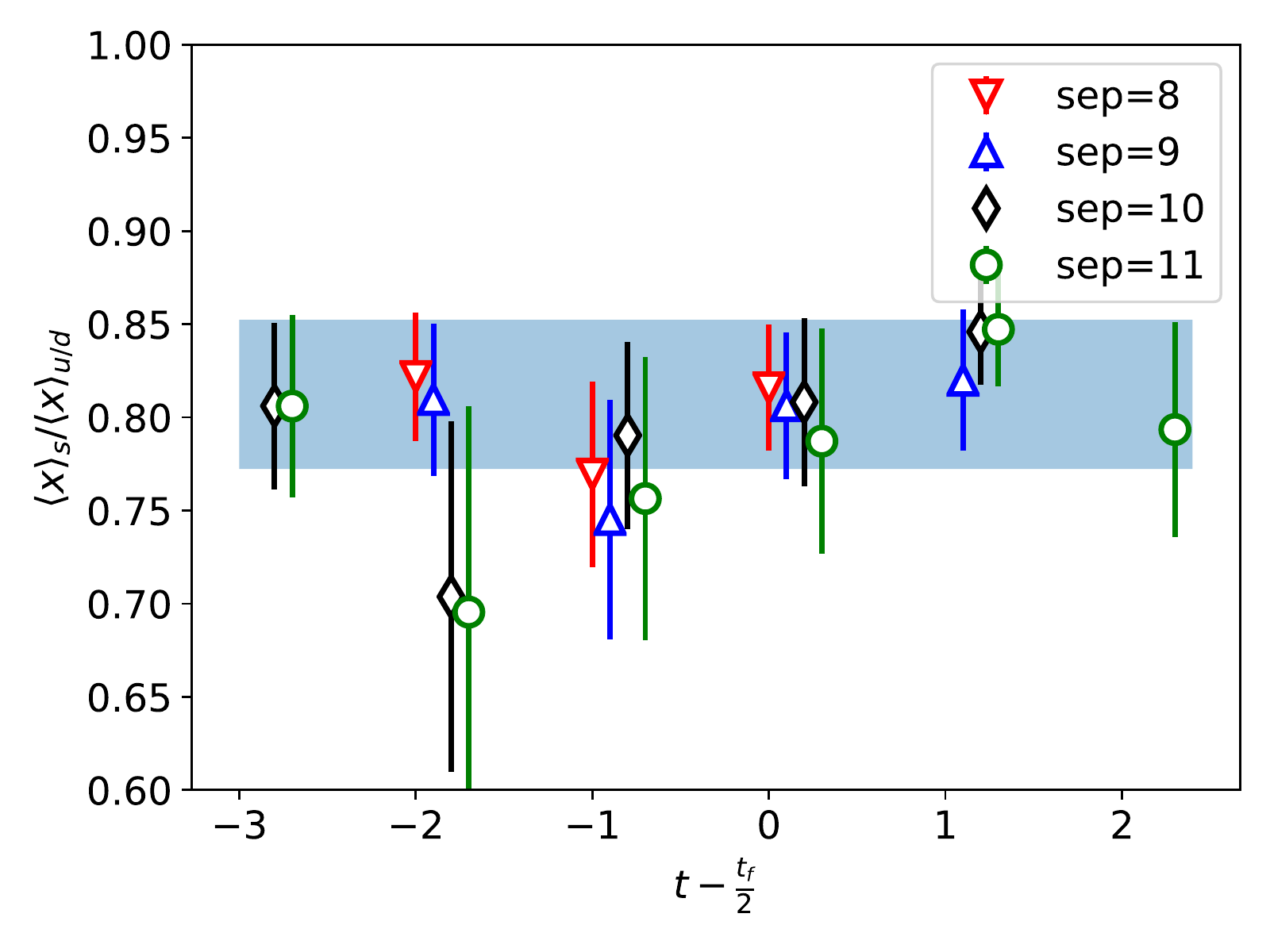}
}
\caption{Example plots of the momentum fractions in DI for the 24I and 32I lattice at their unitary points.
The finite source-sink separation effects are removed from the data points using the results of the two-state fits.
The blue bands show the same two-state fitting values as in Fig.~\ref{DI1}.
\label{fig:DI2}}
\end{figure}

Another way to look at the fitting of the ratios
is plotted in Fig.~\ref{fig:DI2}. There, the data points are modified such that
the finite source-sink separation effects are removed using the results of the two-state fits.
We can see that the new data points at different source-sink separations are all consistent with each other
within errors and they all coincide with the fitted results (the blue bands), which means the
two-state fits can successfully track the excited-state effects.

\begin{figure*}[htbp]
\subfloat[48I, $d$ quark]{
\includegraphics[scale=0.48]{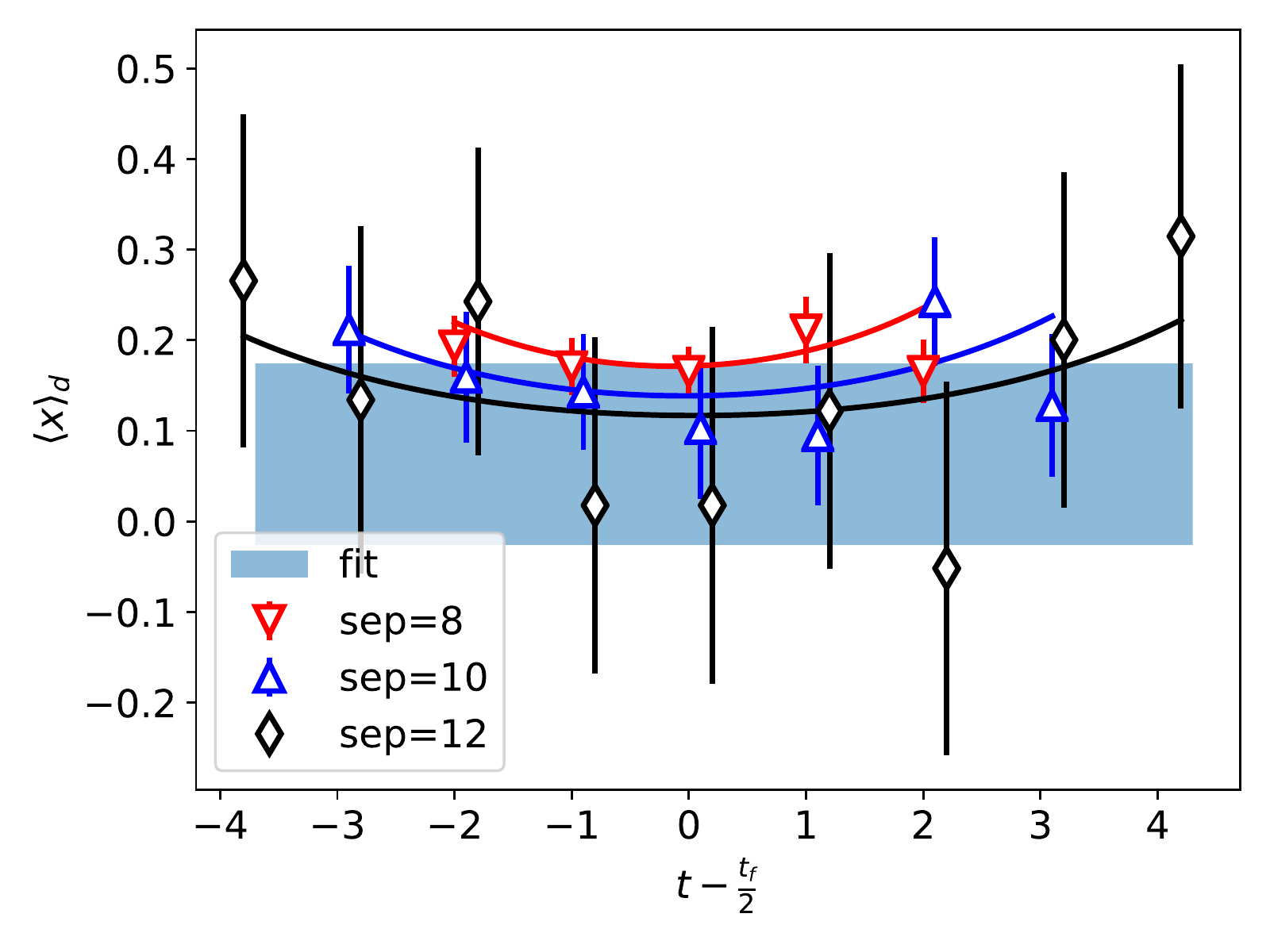}
}
\subfloat[48I, $u$ quark \label{}]{
\includegraphics[scale=0.48]{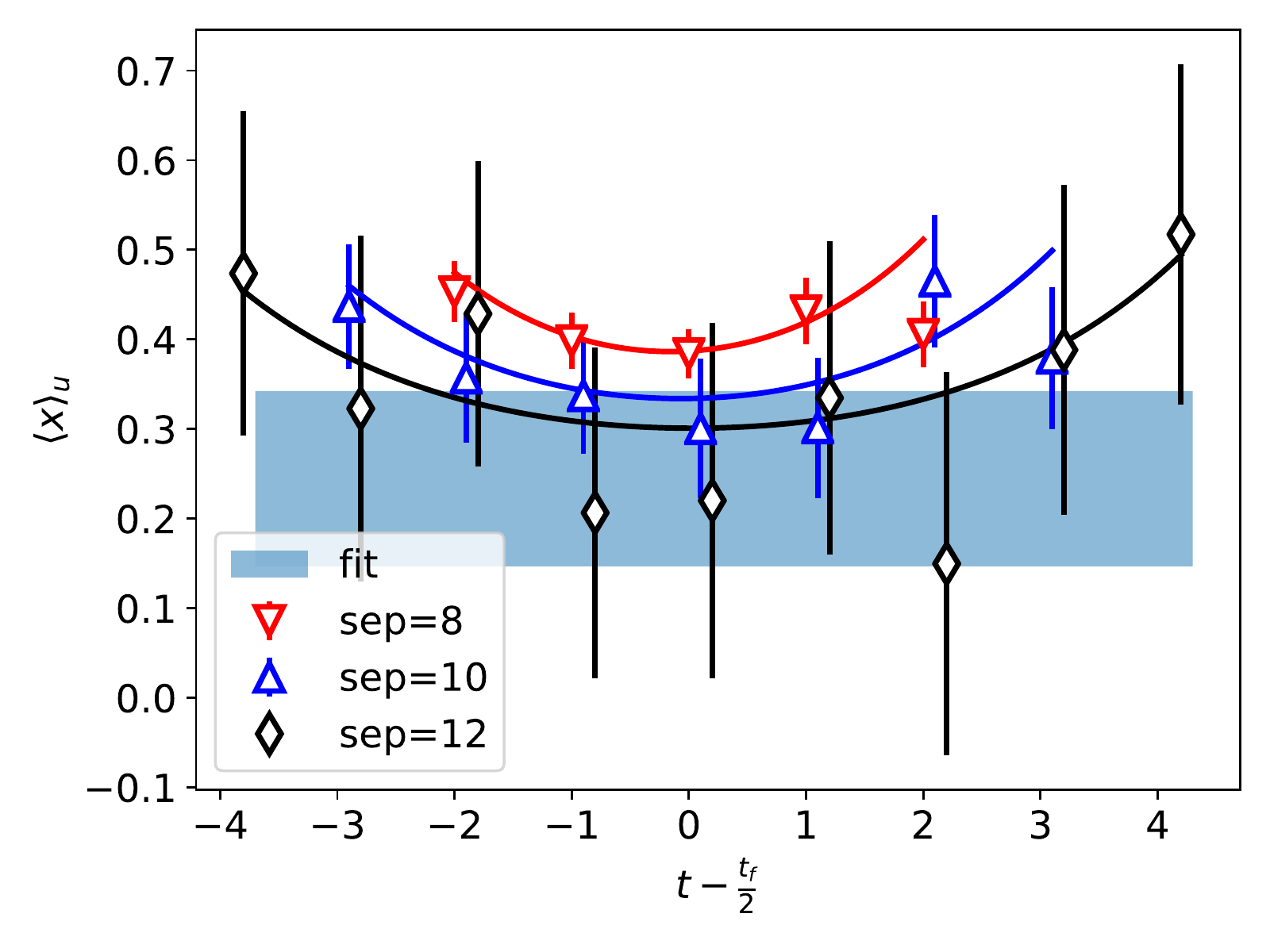}
}\\
\subfloat[24I, $d$ quark\label{}]{
\includegraphics[scale=0.48]{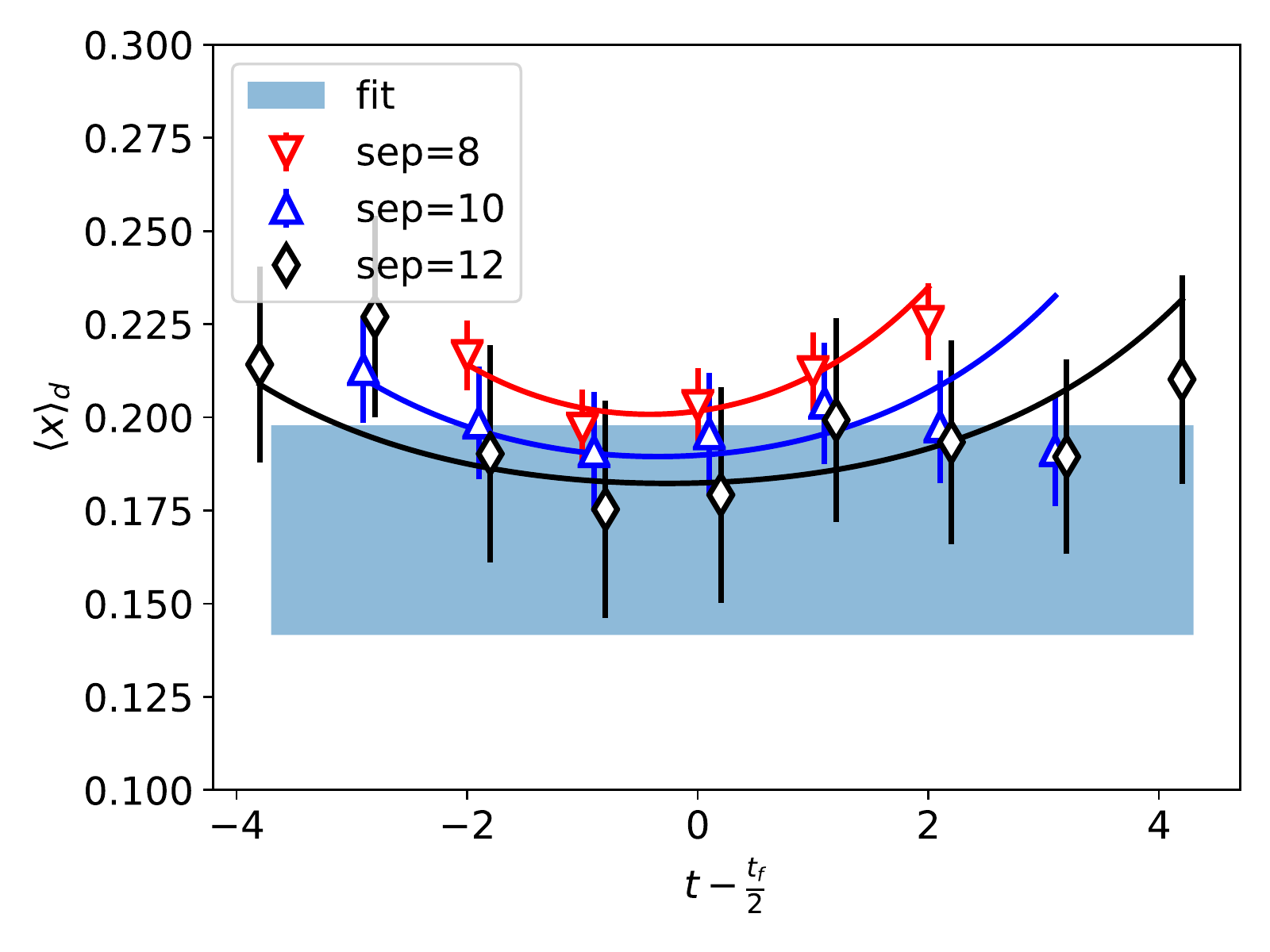}
}
\subfloat[24I $u$ quark\label{}]{
\includegraphics[scale=0.48]{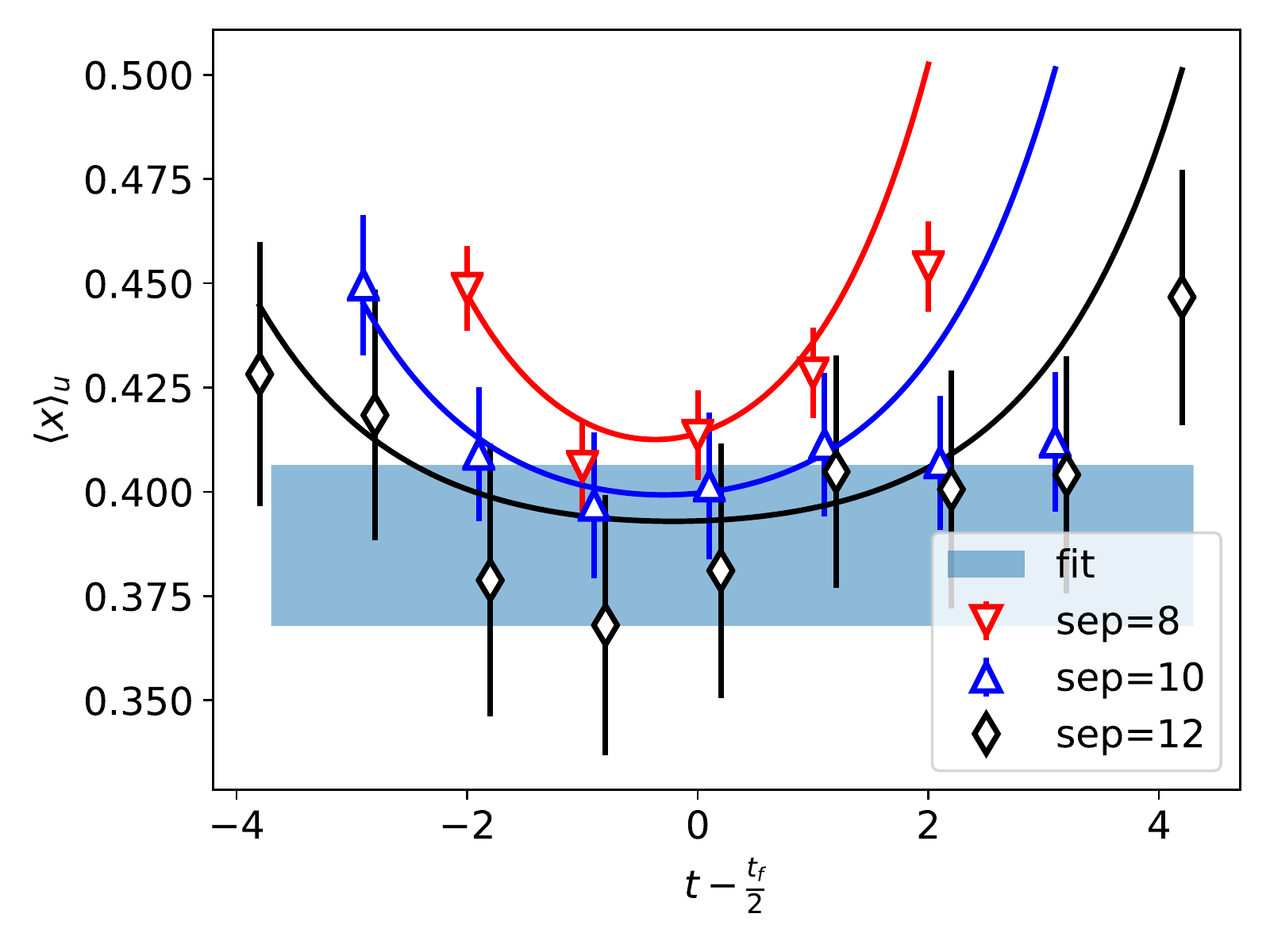}
}
\caption{Similar to Fig.~\ref{DI1}, but for the momentum fractions in CI for the 48I and 24I lattices 
at their unitary points.
$d$ and $u$ quarks are plotted separately. \label{fig:CI}}
\end{figure*}

Although the main topic of this work is to calculate the ratio of DI, the CI part
contributes too through the mixing. 
Example plots of the momentum fractions in CI for the 48I and 24I lattice are shown 
in Fig.~\ref{fig:CI}.
As in the DI case, all the fittings result in good $\chi^2/d.o.f.$ (around or less than one).
Detailed fitting setups can be found in Table~\ref{tab:2s_fitting_CI}.

\begin{table}[htbp]
\caption{Setups of the two-state fits in the CI case.
The source-sink separations used in the fits, the number of points
dropped on the source side and the sink
side and the prior value and width of $\delta m$ are listed for each lattice. \label{tab:2s_fitting_CI}}
\centering{}%
\begin{tabular}{ccccc}
\cline{1-5}
\hline
\multicolumn{1}{c}{lattice} &\multicolumn{1}{c}{separations (a)}&\multicolumn{1}{c}{source drop}
&\multicolumn{1}{c}{sink drop}&\multicolumn{1}{c}{prior $\delta ma$}\tabularnewline
\hline 
32ID/$u$ & 9, 10, 11 & 3 & 2 & 0.4(0.2)\tabularnewline
\hline 
24I/$d$ & 8, 10, 12 & 2 & 1 & 0.3(0.3)\tabularnewline
\hline 
48I/$d$ & 8, 10, 12 & 2 & 1 & 0.3(0.3)\tabularnewline
\hline 
32I/$d$ & 12, 14, 15 & 2 & 3 & 0.2(0.1)\tabularnewline
\hline 
\end{tabular}
\end{table}

After renormalization, the final ${\cal R}$ ratios on different ensembles with different valence pion masses are fitted by the following
form to track the pion mass, lattice spacing and volume dependence
\begin{eqnarray}\label{global}
{\cal R}&&(m_{\pi}^{v},m_{\pi}^{s},a,L)={\cal R}(m_{\pi}^0,m_{\pi}^0,0,\infty)+C_{1}\left((m_{\pi}^{v})^{2}-(m_{\pi}^{0})^{2}\right)\nonumber\\
&&+C_{2}\left((m_{\pi}^{v})^{2}-(m_{\pi}^{s})^{2}\right)+C_{3}^{I/ID}a^{2}+C_{4}e^{-m_{\pi}^{v}L},
\end{eqnarray}
where the $C$'s are free parameters, $m_{\pi}^{v}$/$m_{\pi}^{s}$ is the valence/sea pion mass, and $m_{\pi}^{0}$ is
the physical pion mass. The third term is to account for the partial quenching effect.
A total of 21 data points are used for this global analysis.
The extrapolated result to the physical limit is ${\cal R}^{\overline{\rm MS}} (2\, {\rm GeV}) =\langle x\rangle_{s+\bar{s}}^{R}/\langle x\rangle_{u+\bar{u}}^{R} ({\rm DI})=0.795(79)(77)$ with $\chi^2/d.o.f.=0.16$, where the
first error is the statistical error and the second error 
is the total systematic one.
A complete breakdown of the systematic uncertainties
can be found in detail in Table~\ref{errors}. Details are discussed as follows.

\begin{table*}[htbp]
\caption{The systematic error budget of the ratio ${\cal R}$. \label{errors}}
\centering{}%
\begin{tabular}{c|c|c}
\hline
source & absolute error & relative error\\
\hline
two-state fit (including the use of prior on $\delta m$) & 0.050 & 6.3\% \\

finite lattice spacing & 0.003 & 0.4\% \\
finite lattice volume & 0.027 & 3.4\% \\
pion mass extrapolation & 0.045 & 5.7\% \\

mixed action & 0.016 & 2.0\% \\
strange quark mass & 0.008 & 1.0\% \\

renormalization and mixing & 0.016 & 2.0\% \\
lack of mixing from charm quark & 0.004 & 0.5\% \\
\hline
all  (combined in quadrature) & 0.077 & 9.7\% \\

\hline 
\end{tabular}
\end{table*}
%includes the
%systematic uncertainties from the chiral, continuum, and infinite
%volume interpolation/extrapolations. 
%The dependence on the strange quark mass 
%%in the gauge ensembles and the valence strange quark mass 
%is ignored,
%because it is usually very
%weak as observed in our previous studies of other quantities (e.g., the sigma term~\cite{Yang:2015uis}).

The systematic uncertainty coming from the two-state fits is
estimated in a way shown in Fig.~\ref{fig:DI2}.
If the two-state fits work well,
after the finite source-sink separation effect is removed
from the data points by using the results of the two-state fits,
all the points at different source-sink separations should lie
on the same horizontal line with respect to the
current insertion time.
Fig.~\ref{fig:DI2} shows that this is true within errors.
We then use 
%the difference between the central values of the modified data
%points and 
the difference between the modified data
points and the two-state fitting results 
to estimate the corresponding systematic uncertainty to be $\sim 6\%$ of the central value of
the final result.

\begin{figure}
    \includegraphics[page=1, width=0.45\textwidth]{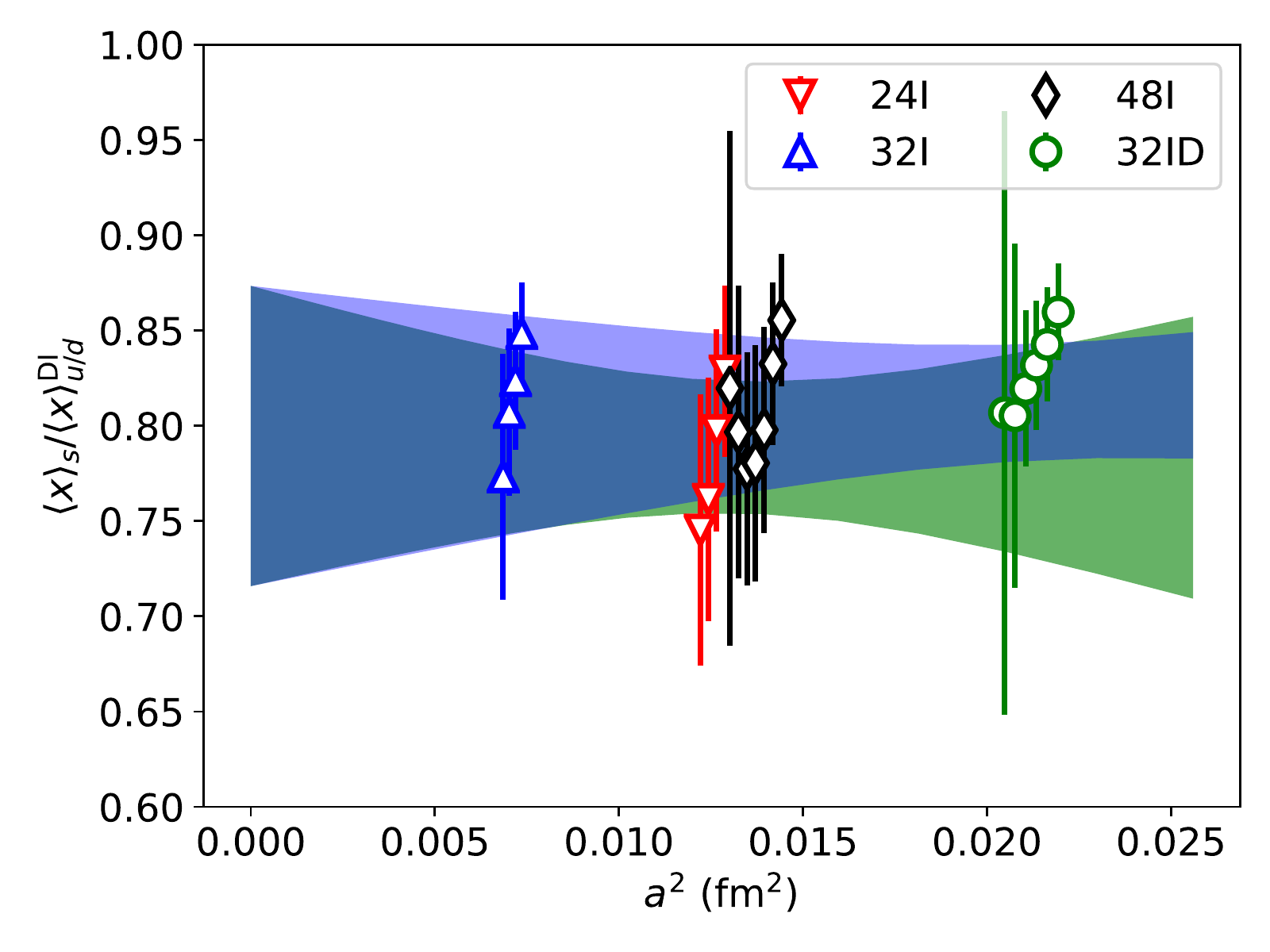}
    \includegraphics[page=1, width=0.45\textwidth]{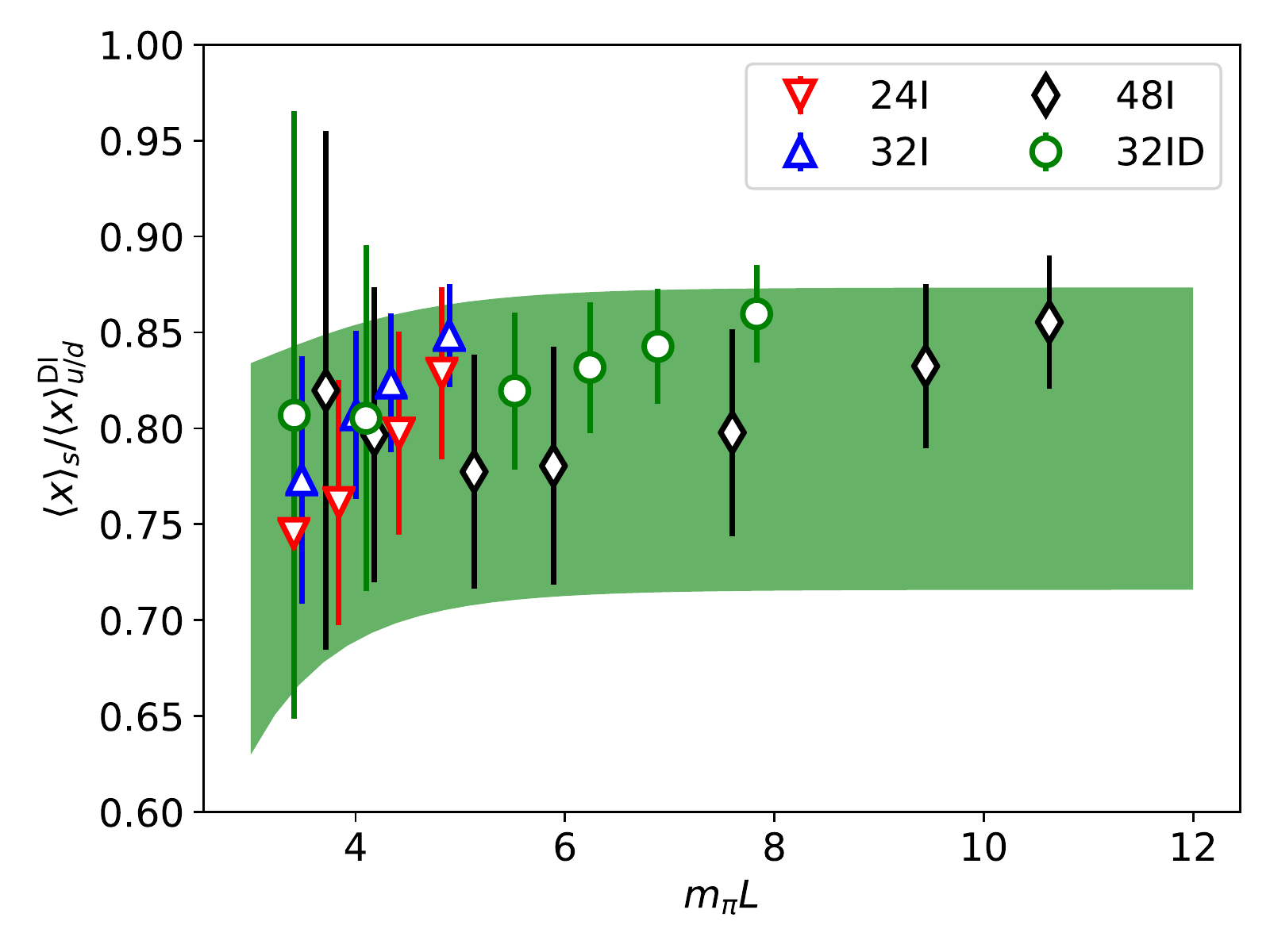}
    \caption{The ${\cal R}$ ratios as a function of lattice spacings (upper panel) and as a 
    function of $m_\pi L$ (lower panel). The
    green band and blue band in the upper panel show the lattice spacing dependence of $\cal R$ at the physical $m_\pi$ and 
    at the infinite volume
    limit for the ``I'' and ``ID'' lattices, respectively. The green band in the lower panel shows the volume dependence 
    of $\cal R$ at the physical $m_\pi$ and at the continuum limit.\label{check}}
\end{figure}

The systematic uncertainties related to the global extrapolation, including 
finite lattice spacing effect, finite lattice volume effect, 
pion mass extrapolation with mixed action and strange quark mass effect,
are estimated following Ref.~\cite{Yang:2015uis}.
We use the difference between
the continuum-extrapolated result and the result on our finest lattice, i.e.\ 32I, at the
physical pion mass point and at the infinite volume limit
for the systematic uncertainty of the finite lattice spacing effect.
We have two coefficients for the finite lattice 
spacing effects in Eq.~(\ref{global}) since we are using two kinds of gauge actions. 
The results, $C_3^I=-0.4(5.3)$ and $C_3^{ID}=0.8(3.3)$, have
no statistical significance for both cases, which is consistent with the behavior our data (the upper panel of Fig.~\ref{check}).
Numerically, the corresponding uncertainty is $|C_3^I|\times a^2({\rm 32I})\sim 0.003$.
It is also consistent with some other calculations 
with the overlap fermion where there are no discernible discretization effects within statistics~\cite{Liang:2018pis}.
The lower panel of Fig.~\ref{check} shows the $\cal R$ dependence on lattice volume.
The band shows the fitted volume dependence at the physical $m_\pi$ and at the continuum limit.
%We see that at $m_\pi L\sim 3$ (48I and 32ID with near physical pion masses), 
%although the error is large, there is visible finite volume effect.
From the band, visible finite volume effect can be seen at $m_\pi L< 4$.
Numerically, the coefficient of the volume effect $C_4$ in Eq.~(\ref{global}) is $-1.3(1.3)$. It has only a one-sigma signal but
is more statistically significant than the coefficients of the lattice spacing effect. Using the same strategy, the corresponding
systematic uncertainty is estimated to be $|C_4|\times e^{-m_{\pi,{\rm phy}} L(\rm{48I})}\sim 0.027$.

Some details about the chiral extrapolation are discussed as follows. 
For a partially quenched chiral extrapolation,
one needs to separate the valence pion mass $m_{\pi,vv}$ and the sea pion mass $m_{\pi,ss}$ dependence.
Additionally, for a mixed action case where the valence is a chiral fermion (as in our case where we use the overlap fermions),
the LO chiral Lagrangian involves 
only one more low-energy constant~\cite{Chen:2006wf} and the correction vanishes as ${\cal O}(a^2)$:
\begin{equation}
    {m'}_{\pi,vs}^2 = \frac{1}{2}\left( m_{\pi,vv}^2 +m^2_{\pi,ss}\right)+a^2\Delta_{\rm mix},
\end{equation}
where the term $a^2\Delta_{\rm mix}$ gauges the difference between 
the valence and sea lattice actions at a finite lattice spacing $a$.
Also, we found that, when using overlap valence on domain wall sea,
$\Delta_{\rm mix}$ is much smaller than those of other mixed action combinations.
For example, 
it is an order of magnitude smaller than that of domain wall on staggered fermion~\cite{Orginos:2007tw,Aubin:2008wk}
and only shifts
$m_{\pi,vs}$ from the would be unitary mass $m_\pi$ by $\sim 10$ MeV for $m_\pi\sim300$ MeV 
at $a=0.11$ fm~\cite{Lujan:2012wg}.
In the chiral extrapolation of the ratio $\cal R$, 
the errors of the data are around 10\% or even larger and one cannot fit complicated pion mass
dependence except for the leading linear terms in $m_{\pi,vv}^2$ and $m_{\pi,ss}^2$,
in which case the delta mix effects are already included in the $a^2$ extrapolation. 
Additional and higher order terms 
(including the ${m'}_{\pi,vs}$ terms appearing in their third power) 
show no statistical significance which
means the difference caused by ignoring higher order terms is much smaller than the current statistical uncertainty.
For the strange quark mass part,
as mentioned before, the bare valence strange quark masses are determined by the global-fit
value at 2 GeV in the $\overline{{\rm MS}}$ scheme calculated in our previous study~\citep{Yang:2014sea}. 
For all the 24I, 48I and 32I lattices the renormalized strange quark mass
is around 100.5 MeV and for the 32ID lattice the number is around 95 MeV, which are all 
consistent with our global-fit
value 101(3)(6) MeV~\citep{Yang:2014sea} within error. As discussed in Ref.~\cite{Yang:2015uis},
the valence strange quark mass effect is very weak and 
we assume so for the sea strange quark too. 
So combining both the valence and sea effects,
we estimate the systematic uncertainty due to strange quark mass to be around 1\%.

The systematic uncertainties corresponding to the renormalization are discussed in detail in the 
supplemental materials of Ref.~\cite{Yang:2018nqn}. They contribute in total $\sim 2\%$
systematic uncertainty of ${\cal R}$.
In addition, we did not consider the charm quark contribution by mixing.
However, the charm momentum fraction itself is small. 
%For example in a recent lattice calculation (2003.08486), 
%the bare $\langle x \rangle_s=0.038(10)$ while $\langle x \rangle_c=0.008(8)$. i
More importantly, 
the sea quark-to-quark mixing coefficients are of order one percent as given in our paper, so the neglect of the heavy flavor
contributions is safe and we estimate that it can only leads to $\sim 0.5\%$ uncertainty. 

We use the traceless
diagonal part of the EMT $\bar{T}_{44}$ to carry out our calculation.
Actually, in our previous papers discussing the glue momentum fraction renormalization~\cite{Yang:2018bft}
and nucleon mass decomposition~\cite{Yang:2018nqn},
we found that the assumed rotational symmetry breaking effect of the renormalization constant
between the off-diagonal EMT $T_{4i}$
and the traceless
diagonal part $\bar{T}_{44}$ which are in different irreducible representations of the cubic point group $O_h$
is much smaller than the statistical uncertainties for both the glue and quark case.
So we do not add one more systematic uncertainty due to the rotational symmetry breaking.
In total, as shown in the table, the systematic uncertainty 
of ${\cal R}$ is about 9.7\% which is close to the statistical uncertainty.

\begin{figure}[htbp]
\centering{}\includegraphics[scale=0.45]{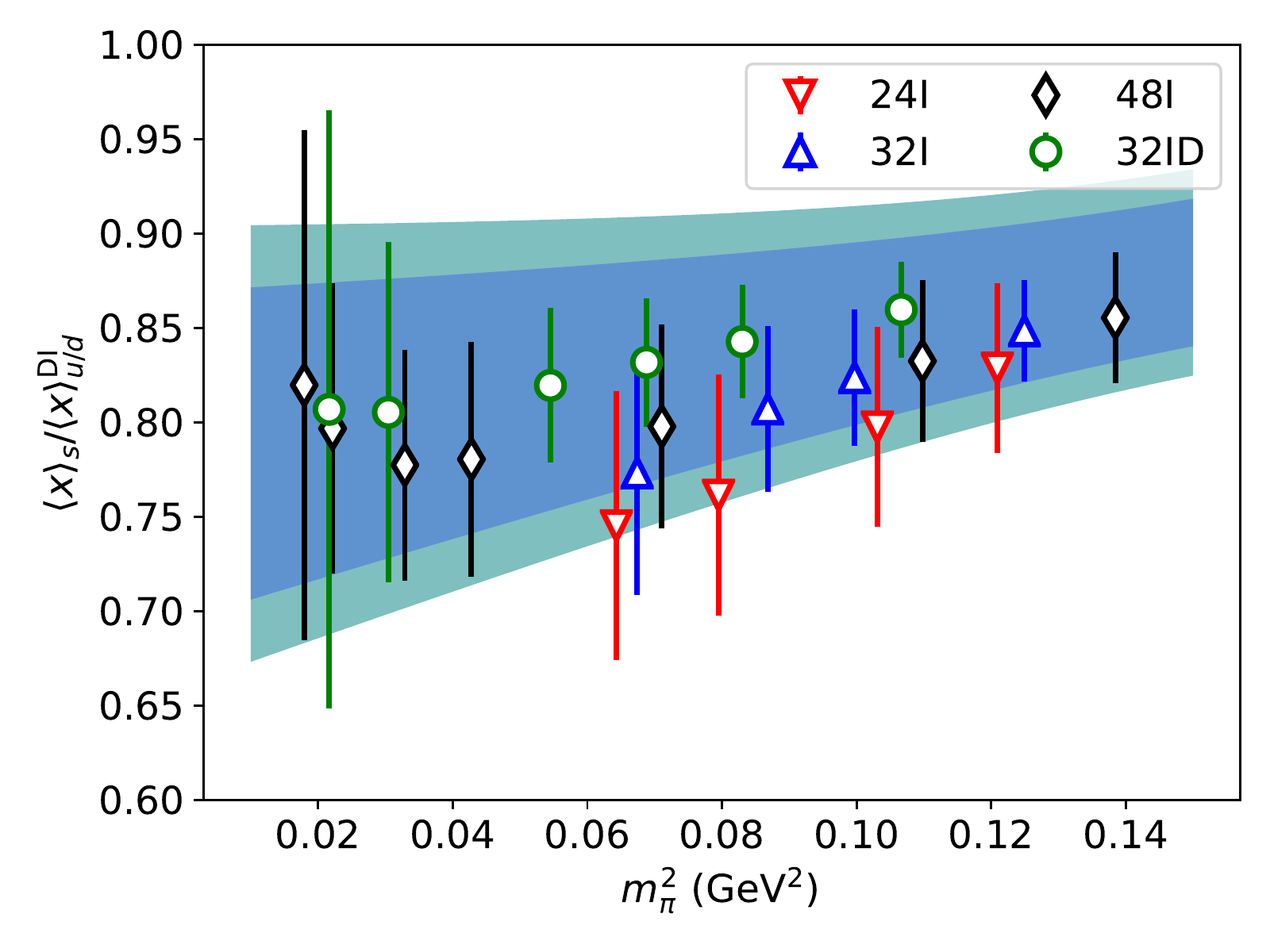}
\caption{The global interpolation/extrapolation on the four ensembles. The
blue and cyan bands show the statistical and total uncertainties of
our final prediction. \label{fig:dis-ratio}}
\end{figure}

The bands in Fig.~\ref{fig:dis-ratio} show our final prediction on the unitary pion mass dependence of the ratio in the continuum and infinite volume limits.
The width of the blue band indicates the statistical error and the width of the wider cyan band the total error.
The data points from different ensembles with the partially quenching effect subtracted are also plotted in the figure.
%The agreement between the data points and the bands shows that the finite volume and lattice spacing effects are small.
The agreement of the bands and the data points shows that the finite volume and lattice spacing effects are small.

\section{Discussion and Summary}

\begin{figure}[htbp]
\begin{centering}
%\vspace{5pt}
\includegraphics[scale=0.45]{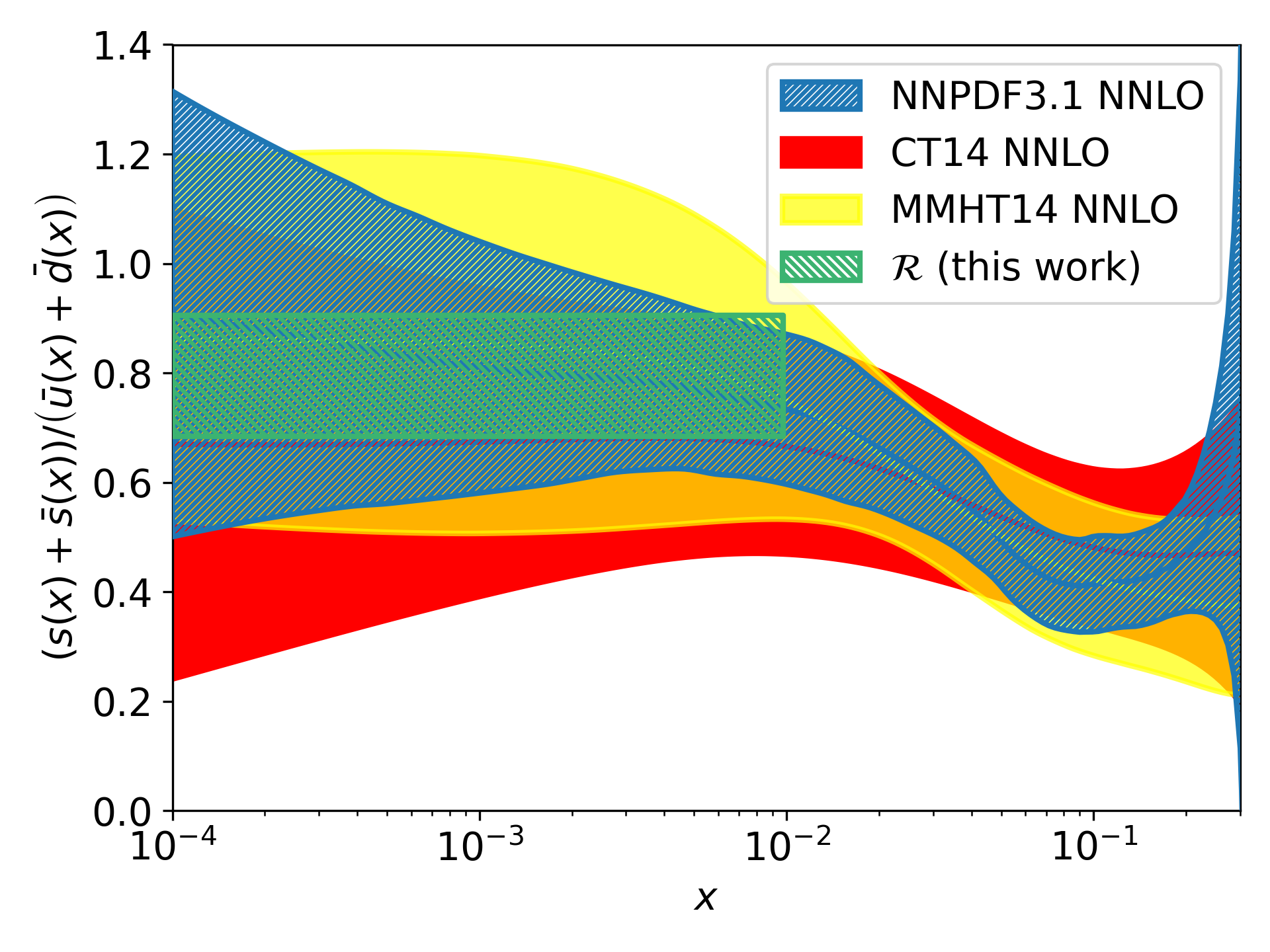}
%\vspace{30pt}
\par\end{centering}
\centering{}\caption{The global fitting results of $\left(s(x)+\bar{s}(x)\right)/\left(\bar{u}(x)+\bar{d}(x)\right)$ at $Q^2=4$ GeV$^2$. The green band shows our result under the assumption that 
the ratio is a constant for small $x$ up to $x = 10^{-2}$. \label{fig:PDFs}}
\end{figure}

%To show another manifestation of the existence of the CS degrees of freedom besides the explicit evidence from
%the Gottfried sum rule violation,
To manifest that the CS and DS partons have distinct small $x$ behaviors,
%To see how the lattice $\cal R$ value helps the global PDF analysis, we show in Fig.~\ref{fig:PDFs}, 
the ratio of the distribution 
$(s(x)+\bar{s}(x))/(\bar{u}(x) + \bar{d}(x))$ from 3 global fittings at NNLO~\cite{Ball:2017nwa,Dulat:2015mca,Harland-Lang:2014zoa} at $Q^2= 4\, {\rm GeV^2}$ is plotted in Fig.~\ref{fig:PDFs}. 
We see that the ratios are kind of flat
at small $x \lesssim 10^{-2}$ with large errors.
On the other hand,
it is conspicuous that they all have a
characteristic shoulder with a fall off around $x\sim 10^{-2}$ toward larger $x$. 
Since under our classification, $s(x)+\bar{s}(x)$ contains only the DS contribution while $\bar{u}(x) + \bar{d}(x)$ includes both CS and DS, 
this reflects the fact that the small $x$ behavior of $\bar{q}^{ds}(x)$ 
is more singular than that of $q^v$ and $\bar{q}^{cs}$ 
such that at $x \lesssim 10^{-2}$ the DS dominates, so
the ratio stays roughly constant~\cite{Liu:1999ak,Liu:2012ch}.
%When $x$ approaches $10^{-2}$ from below, 
When $x$ is larger than $\sim 10^{-2}$, 
the CS $\bar{u}^{cs}(x) + \bar{d}^{cs}(x)$ component
in $\bar{u}(x) + \bar{d}(x)$ ({\it N.B.} $\bar{u}(x) + \bar{d}(x) = \bar{u}^{cs}(x) + \bar{d}^{cs}(x) + \bar{u}^{ds}(x) + \bar{d}^{ds}(x)$) sets in to make the ratio smaller.
%This can also be understood in Regge theory where $q^{ds},\bar{q}^{ds}_{\stackrel{\longrightarrow}{x \rightarrow 0}} x^{-1}$ while 
%$q^v, \bar{q}^{cs}_{\stackrel{\longrightarrow}{x \rightarrow 0}} x^{-1/2}$.
This can also be understood in Regge theory 
where $q^v, \bar{q}^{cs}_{\stackrel{\longrightarrow}{x \rightarrow 0}} x^{-1/2}$ since
the CS partons are in the connected insertion which is flavor non-singlet as are the valence partons
and their small $x$ behaviors reflect the leading Reggeon exchanges~\cite{Brodsky:1990gn}. 
On the other hand,
$q^{ds},\bar{q}^{ds}_{\stackrel{\longrightarrow}{x \rightarrow 0}} x^{-1}$ since
the DS is flavor singlet
and can have Pomeron exchanges~\cite{Kuti:1971ph,Reya:1979zk}.
Also, we find numerically that
in the small $x$ region ($10^{-4}$ to $10^{-2}$) of the global fittings for PDF~\cite{Ball:2017nwa,Dulat:2015mca,Harland-Lang:2014zoa,Buckley:2014ana} at $Q^2 = 4$ GeV$^2$, 
the power of the small $x$ behavior 
%%(i.e. the $\alpha$ in $x^{\alpha}$) 
for $\bar{q}(x)$ is in the range $[-1.22, -1.15]$ and for $q^v(x)$ in the range [-0.6, -0.2], which are close to those prescribed in Regge theory and 
consistent with our argument that DS dominates the small $x$ behavior.
%This behavior reflects the fact that the small $x$ behavior of $\bar{q}^{ds}(x)$ 
%is more singular than that of $q^v$ and $\bar{q}^{cs}$ (e.g. $q^{ds},\bar{q}^{ds}_{\stackrel{\longrightarrow}{x \rightarrow 0}} x^{-1}$ and 
%$q^v, \bar{q}^{cs}_{\stackrel{\longrightarrow}{x \rightarrow 0}} x^{-1/2}$ in Regge theory and
%for the small $x$ region ($10^{-4}$ to $10^{-2}$) of the global fittings of PDF~\cite{Ball:2017nwa,Dulat:2015mca,Harland-Lang:2014zoa} at $Q^2 = 4$ GeV$^2$, 
%we find the power of the small x behavior (i.e. $\alpha$ in $x^{\alpha}$) for $\bar{q}(x)$ is in the range $[-1.22, -1.15]$, and $q^v(x)$ in the range [-0.6, -0.2] which are close to those prescribed in Regge theory)
%so that at $x < 10^{-2}$, where the DS dominates,
%the ratio stays more or less constant. 

The dominance of the DS partons at small $x$ reveals the possibility that lattice QCD can help to constrain
the strange quark distribution in the small $x$ region using the DI ratio $\cal R$
as defined in Eq.~(\ref{Eq:R}).
%because the DI ratio approaches the DS ratio
%at small $x$ and the ratio can be determined precisely on the lattice.
%These behaviors serve as an indirect manifestation of the existence of the CS degrees of freedom besides the explicit evidence from
%the Gottfried sum rule violation.
To do so,
%In order to reveal these new parton degrees of freedom and the physics hidden behind them, 
it is advocated~\cite{Liu:2017lpe} to 
completely separate the CS and DS partons with their corresponding evolutions in new global fittings.
%where 
%definitive lattice calculations like the $\cal R$ in this work 
%(and possibly other lattice moments) can be used as constraints 
%to pin down the PDFs at small $x$.
%together with experiments.}
%with both statistical and systematic uncertainties under control
%should be treated equally to experiments.}
%the $\cal R$ value from lattice
%calculation (and possibly other lattice moments) can be used as a constraint
%{\color{red}to pin down the PDFs at small $x$.
%Definitive lattice calculations like the $\cal R$ in this work 
%with both statistical and systematic uncertainties under control
%should be treated equally to experiments.}
No approximation such as 
$u^{ds}(x)=\bar{u}^{ds}(x)$
is needed in this approach. Once done, close connection can be built between lattice computations 
of moments and those from global fittings.

In addition, before new global fittings that separate CS and DS are carried out, %%which we understand could be complicated and time-consuming, 
an easy-to-implement first trial is to use the following approximate form for the strange PDF
\begin{equation}
\label{proposal2}
s(x)+\bar{s}(x) = \frac{1}{\cal R}\left(\bar{u}(x) + \bar{d}(x)\right)  - c \left(\bar{u}(x) - \bar{d}(x)\right),\\
%\bar{u}(x) + \bar{d}(x) = {\cal R} \left(s(x)+\bar{s}(x)\right) + c \left(\bar{u}(x) - \bar{d}(x)\right),
\end{equation}
where $c$ is a coefficient to be fixed, to better control the statistical uncertainty of the global fittings.
%%The form also comes from the fact that $s(x)+\bar{s}(x)$ represents the DS contribution which dominates the 
%%small $x$ behavior of $\bar{u}(x) + \bar{d}(x)$ while $\bar{u}(x) - \bar{d}(x)$ represents the CS contribution which dominates its
%%arge $x$ behavior.
The form is based on the approximation that $s(x)+\bar{s}(x)$ is proportional to $\bar{u}^{ds}(x)+\bar{d}^{ds}(x)$
by the factor of $\cal R$ if $u^{ds}(x)=\bar{u}^{ds}(x)$ and isospin symmetry are assumed.
Since the first term $\bar{u}(x)+\bar{d}(x)$ in Eq.~(\ref{proposal2}) contains both $\bar{u}^{ds}(x)+\bar{d}^{ds}(x)$ and $\bar{u}^{cs}(x)+\bar{d}^{cs}(x)$,
%the latter needs to be subtracted from the first term of Eq.~(\ref{proposal2}).
we use the second term to subtract the latter which
%It 
is taken to be proportional to $\bar{u}(x)-\bar{d}(x)$ due to the fact that it equals $\bar{u}^{cs}(x)-\bar{d}^{cs}(x)$ in the isospin limit.
%%Thus the second term in Eq.~(\ref{proposal2}).
Both $\bar{u}(x)+\bar{d}(x)$ and $\bar{u}(x)-\bar{d}(x)$ in Eq.~(\ref{proposal2}) are obtained in global fittings, 
thus this form should be easy to implement.
This form serves as an explicit example of how a lattice result enters directly the global fittings of PDFs.
%The approximation of $\bar{u}=\bar{d}$ is assumed to plug in the lattice calculation of $\cal R$.
Further lattice calculation of the fourth moment $\langle x^3\rangle$ of the DI will serve to gauge the validity of this approach and suggest
possible modification of the fitting function.
We also plot our $\cal R$ in Fig.~\ref{fig:PDFs} up to $x = 10^{-2}$ to visually show that
the uncertainty of the strange PDF can be reduced significantly by using this lattice constraint as indicated by the
lattice error as compared to those from the NNLO analyses. 

%To visually show how the statistical uncertainty can be reduced by using this lattice constraint compared with the present results from global analyses of experiments, 
%we also plot our $\cal R$ in Fig.~\ref{fig:PDFs} up to $x = 10^{-2}$.
%As pointed out in the above discussion,
%DS dominates $\bar{u}(x) + \bar{d}(x)$ in the small $x$ region.
%So to the extent that $s(x)+\bar{s}(x)$ and $\bar{u}^{ds}(x) + \bar{d}^{ds}(x)$ are largely proportional, 
%the ratio $\cal R$ would represent the PDF ratio
%$(s(x)+\bar{s}(x))/(\bar{u}(x) + \bar{d}(x))$ at small $x$ up to higher moments corrections. 
%This shows that if the lattice $\cal R$ is used as a constraint in a global analysis, the statistical uncertainty of the strange PDF can be reduced significantly.
%as indicated by
%its error as compared to those from the NNLO analyses. 

%%Besides the hadronic tensor~\citep{Liu:1999ak,Liu:2016djw,Liang:2019frk}, recent formalisms~\citep{Ji:2013dva,Chambers:2017dov,Radyushkin:2017cyf,Ma:2017pxb},  
%%have been developed to calculate the explicit $x$-dependent PDF on the lattice. But it is still a challenge for these approaches to have all the statistics and systematics under control at this stage. 
Since
lattice calculations with low quark and glue moments are getting mature and complete,
%with nonperturbative renormalization
%and mixing taken into account 
the
QCD path-integral
classification that separates CS and DS extends the ability that lattice calculations can 
serve as meaningful constraints for the global analysis of PDFs. 
The present result of the ratio
$\langle x\rangle_{s+\bar{s}}/\langle x\rangle_{u+\bar{u}} ({\rm DI})=0.795(79)(77)$ at $\overline{\rm MS}$ scale $\mu = 2$ GeV is
the first such calculation that can constrain the global fittings in the small $x$ region. Our results will have important impact 
on future global fittings and experiments 
%such as EicC~\cite{Chen:2019equ} which focuses on the sea quark physics.
from EIC and LHC.

\begin{acknowledgments}
We thank the RBC and UKQCD Collaborations for providing their DWF
gauge configurations. This work is supported in part by the U.S. DOE
Grant No. DE-SC0013065 and DOE Grant No. DE-AC05-06OR23177 which is within the framework of the TMD Topical Collaboration.
%Y.Y is supported in part by Chinese Academy of Science CAS Pioneer Hundred Talents Program.
This research used resources of the Oak Ridge
Leadership Computing Facility at the Oak Ridge National Laboratory,
which is supported by the Office of Science of the U.S. Department
of Energy under Contract No. DE-AC05-00OR22725. This work used Stampede
time under the Extreme Science and Engineering Discovery Environment
(XSEDE), which is supported by National Science Foundation Grant No.
ACI-1053575. We also thank the National Energy Research Scientific
Computing Center (NERSC) for providing HPC resources that have contributed
to the research results reported within this paper. We acknowledge
the facilities of the USQCD Collaboration used for this research in
part, which are funded by the Office of Science of the U.S. Department
of Energy.
\end{acknowledgments}

\bibliographystyle{apsrev4-1}
\bibliography{library}

\end{document}